# Multiplex stimulated Raman scattering imaging cytometry reveals cancer metabolic signatures in a spatially, temporally, and spectrally resolved manner

Kai-Chih Huang[1,†]; Junjie Li[2,†]; Chi Zhang[2,†]; Yuying Tan[1], and Ji-Xin Cheng[1,2,3,*]

[1] Department of Biomedical Engineering, Boston University, Boston, MA 02215, USA.
[2] Department of Electrical and Computer Engineuter Engineering, Boston University, Boston, MA 02215, USA.
[3] Photonics Center, Boston University, Boston, MA 02215, USA.
†These authors contributed equally to this work
Corresponding author e-mail address: jxcheng@bu.edu

***In situ*** **measurement of cellular metabolites is still a challenge in biology. Conventional methods, such as mass spectrometry or fluorescence microscopy, would either destruct the sample or introduce strong perturbations to the functions of target molecules. Here, we present multiplex stimulated Raman scattering (SRS) imaging cytometry as a label-free single-cell analysis platform with chemical specifity, and high-throughput capabilities. Cellular compartments such as lipid droplets, endoplasmic reticulum, and nuclei are seperated from the cytoplasm. Based on these chemical segmentations, 260 features from both morphology and molecular composition were generated and analyzed for each cell. Using SRS imaging cytometry, we studied the metabolic responses of human pancreatic cancer cells under stress by starvation and chemotherapy drug treatments. We unveiled lipid-facilitated protrusion as a metabolic marker for stress-resistant cancer cells through statistical analysis of thousands of cells. Our findings also demonstrate the potential of targeting lipid metabolism for selective treatment of starvation-resistant and chemotherapy-resistant cancers. These results highlight our SRS imaging cytometry as a powerful label-free tool for biological discoveries with a high-throughput, high-content capacity.**

Altered cell metabolism is recognized as one of the hallmarks of cancer (Hanahan and Weinberg, 2011). The reprogrammed metabolism, which is deployed by cancer cells to fulfill the demands of fast proliferation (Heiden et al., 2009; Warburg, 1956), offers new opportunities for diagnosis and treatment of cancer (Pelicano et al., 2006; Hamanaka and Chandel, 2012; Zhao et al., 2013). However, challenges remain in the quantification of cell metabolism, one of which is the inter- or intratumoral heterogeneous metabolic profiles of cancer cells. Although the ensemble measurement of large number of cancer cells identifies specific metabolic features under certain condition, individual cancer cells might show significantly different metabolic response (Altschuler and Wu, 2010). This cell heterogeneity is considered one of the major causes of incomplete tumor remission and relapse (Marusyk and Polyak, 2010; Meacham and Morrison, 2013). Such an inhomogeneous response of cells to environments and drugs cannot be addressed using conventional methods, such as biochemical assays or mass spectrometry, which are based on ensemble measurement of a cell population. High-efficiency imaging tools which can quantify metabolic features of individual cells with subcellular information for a large number of cells become essential.

Flow cytometry is one of the commonly used technologies for high-throughput single-cell analysis, which generates statistical information of a cell population (Hulett et al., 1969).

However, conventional flow cytometry integrates the whole-cell signal to a single intensity measurement, lacking the information in spatial distributions of cellular compounds which are essential to understand cell heterogeneity in many cases. To take spatial information into account, one must switch to imaging cytometry or imaging flow cytometry (Blasi et al., 2016; Henriksen et al., 2011; McGrath et al., 2008; Nitta et al., 2018). The conventional imaging cytometry relies on fluorescence labeling, which is unsuitable to track endogenous metabolites in living cells. Photobleaching, non-specific binding, cytotoxicity, and molecular functional perturbation all raise issues for fluorescence-based measurement (Jensen, 2012; Zanetti-Domingues et al., 2013). In addition, high-specificity fluorescent probes do not exist for most small-molecule metabolites. In order to trace specific metabolic molecules in living cells, molecular analog with fluorescent conjugation might be used (Solanko et al., 2013; Yamada et al., 2007). But the relatively large size of fluorescent probes, when tagged to small metabolic molecules, would significantly alter the bio-functions of these molecules (Lee et al., 2015; Wei et al., 2014).

Raman spectroscopy and microscopy have shown great potentials to solve problems encountered by fluorescence-based imaging. It circumvents the process of labeling and contains rich chemical information to characterize small molecules. Capacity of Raman spectroscopy/microscopy has been demonstrated to discover cholesteryl ester accumulation as a marker for aggressive cancer cells (Yue et al., 2014), visualize cytochrome c releases during apoptosis (Okada et al., 2012), and monitor the cellular stage such as macrophage activation (Pavillon et al., 2018). However, spontaneous Raman scattering is a very weak process, resulting in hours to acquire a cellular image, which is impractical for live-cell imaging and imaging cytometry (Zhang et al., 2015a, 2015b). The advent of coherent Raman scattering (CRS) techniques, including coherent anti-Stokes Raman scattering (CARS) and stimulated Raman scattering (SRS), enhanced the Raman efficiency by around seven orders of magnitude (Min et al., 2011; Cheng and Xie, 2016), and achieved imaging speed as fast as fluorescent microscopy (Evans et al., 2005; Ozeki et al., 2012; Saar et al., 2010). CRS microscopy has been used to study lipid metabolism (Fu et al., 2014; Yu et al., 2014, Yue et al., 2014; Li et al., 2017), protein metabolism (Wei et al., 2013, 2014), nucleic acids metabolism(Chen et al., 2014; Wei et al., 2014), retinoid metabolism (Chen et al., 2018; Liao et al., 2015a), cholesterol metabolism (García et al., 2015; Wang et al., 2013; Lee et al., 2015), glucose metabolism (Li and Cheng, 2014; Hu et al., 2015; Zhang et al., 2019), and to monitor small molecular drug delivery (Gaschler et al., 2018; Tipping et al., 2016). To promote high-throughput analysis of single-cells at a high speed, CARS and SRS flow cytometry have been demonstrated (Charles H. Camp et al., 2011; Hiramatsu et al., 2019; Zhang et al., 2017). However, to generate enough signal, CRS usually requires tight laser focusing to a spot much smaller than a cell (Charles H. Camp et al., 2011; Hiramatsu et al., 2019; Zhang et al., 2017). Therefore, CRS signals in flow cytometry might not represent the entire cell. In order to acquire spatial information from the cells and in flow cytometry settings, 4-color SRS imaging flow cytometry was demonstrated recently to classify microalgal cells and cancer cells without the need for fluorescent labeling (Suzuki et al., 2019). However, no subcellular information was obtained and using only 4 colors limits its capability to identify unknown species from complex environment. Despite these advancements, an urgent need for high-throughput *in situ* analysis of cellular metabolism with subcellular information remains unfulfilled.

Here, we designed and constructed a prototype of high-content high-throughput imaging cytometer based on multiplex SRS. The multiplex SRS spectroscopy enabled acquisition of a Raman spectrum covering 200 wavenumbers at a speed of 5 μs in 32 spectral channels. We implemented a hybrid scanning scheme for high-speed hyperspectral cell imaging at a throughput of 30~50 cells per second at diffraction-limited spatial resolution. At a spectral resolution of 13.4 cm$^{-1}$, we segregated the subcellular compartments based on their compositional difference. The high spatial and spectral resolution enables high-content single-cell analysis to address cellular heterogeneity by using our imaging cytometer. Through development of a quantitative analysis algorithm based on CellProfiler, we are able to distinguish 260 morphological and metabolic features in thousands of individual cells, which is not achievable with other technologies. Using our multiplex SRS imaging cytometer, we studied how human cancer cells reprogram their metabolism in response to stress conditions, including starvation and chemotherapy treatment. We found that nutrient starvation or chemotherapy treatment cause lipid droplet redistributions through forming lipid-droplet-facilitated protrusions, which may promote cancer cell survival under stress by enhancing their nutrient uptake capacity. This finding not only opens new opportunities of targeting the reprogrammed lipid metabolic pathway to treat stress-resistant cancer cells but also demonstrates the prowess of multiplex SRS imaging cytometry for discovering important metabolic markers of human diseases.

## Results

**Multiplex SRS-based Label-free Chemical Imaging Cytometry.** To quantify molecular information of large numbers of cells at a high-throughput capacity, we developed a multiplex SRS imaging cytometer. The setup of our multiplex SRS imaging cytometer is shown in **Figure A**. A broadband pump and a narrowband Stokes laser beams simultaneously excite multiple Raman transition modes (**Figure 1B**). After the sample, the pump beam was dispersed by a grating pair and detected by a lab-built 32-channel lock-in free resonant photodiode array detector shown in **Figure 1C**. This parallel-detection scheme allows high-speed, distortion-free, and sensitive Raman spectroscopy at 5 μs per spectrum. At such a speed, we are able to detect dimethyl sulfoxide (DMSO) as low as 0.1% (vol/vol) (**Figure S1A**) covering a 200 cm$^{-1}$ spectral window with a 13.4 cm$^{-1}$ spectral resolution. (**Figure S1B**). A hybrid high-speed scanning was applied by combing one-dimensional laser scanning with a perpendicular one-dimensional translation of motorized stage (**Figure 1A**). With these implantations, our cytometric system achieved the throughput of mapping 30~50 cells per second with diffraction-limited spatial resolution.

**The workflow of single-cell analysis by multiplex SRS imaging cytometry.** We developed a workflow to achieve high-content single-cell analysis using the three-dimensional dataset generated by the high-speed multiplex SRS imaging cytometer. First, *in situ* spectroscopic imaging of a large number of cells is performed (**Figures 1A-1D**). In order to resolve weak chemical compositions, we apply a space-wavelength total variation denoising algorithm to improve the signal-to-noise ratio and image quality (Liao et al., 2015b) (**Figure S2**). Second, chemical maps of sub-cellular compartments including endoplasmic reticulum (ER), lipid droplet (LD), nucleus, and cytoplasm are generated, (**Figure 1D-E**). Here, spectral phasor analysis is utilized to project the SRS spectrum from each pixel (**Figure 1D**) onto a two-dimensional phasor domain (Fu and Xie, 2014) (**Figure 1E**), followed by an unsupervised algorithm to cluster each cellular compartment based on their signature Raman spectrum (**Figure 1E**, details can be found in Methods). Then, we remap the clustered pixels from the phasor domain back to the spatial domain to generate the chemical maps (**Figure 1F**). To validate the segmentation of cellular compartments, we compared our label-free imaging results with the images from fluorescent labeling. Our label-free chemical maps of ER, LD, and nucleus showed good agreement with the fluorescence images of ER, LD, and nucleus labeled by ER tracker, C12-BODIPY, and 4′,6-diamidino-2-phenylindole (DAPI), respectively

(**Figure S3**). Third, to enable high-content single-cell analysis, we used CellProfiler for automatic single-cell segmentation (**Figure 1G**) and feature extraction (**Figure 1H**). A total of 260 features were extracted in each cell to describe cellular or subcellular characteristics. These features were categorized into three groups as shown in **Figure 1H**, including (1) the morphological features, i.e. cell shape, size of ER, nuclear-cytoplasmic ratio, location of LD, etc; (2) SRS intensity features. SRS signal is linearly proportional to the molecular concentration. Therefore, the average intensity reflects molecular concentration, the integrated intensity designates the total amount of chemicals, and the intensity distribution encloses texture information; (3) SRS spectral features. These features show chemical compositions of certain compartments in cells. Specifically, we measured and compared ratio of triglyceride and cholesterol contents in LDs. For each feature, it can be displayed as a histogram (**Figure 1H**) to reflect the distribution of this feature in a cell population. As an example shown in **Figure 1H**, major axial length histogram illustrates the cellular morphology distribution. We confirm the consistency of quantified major axes value by validating the corresponding images of long, medium, and short major axes value.

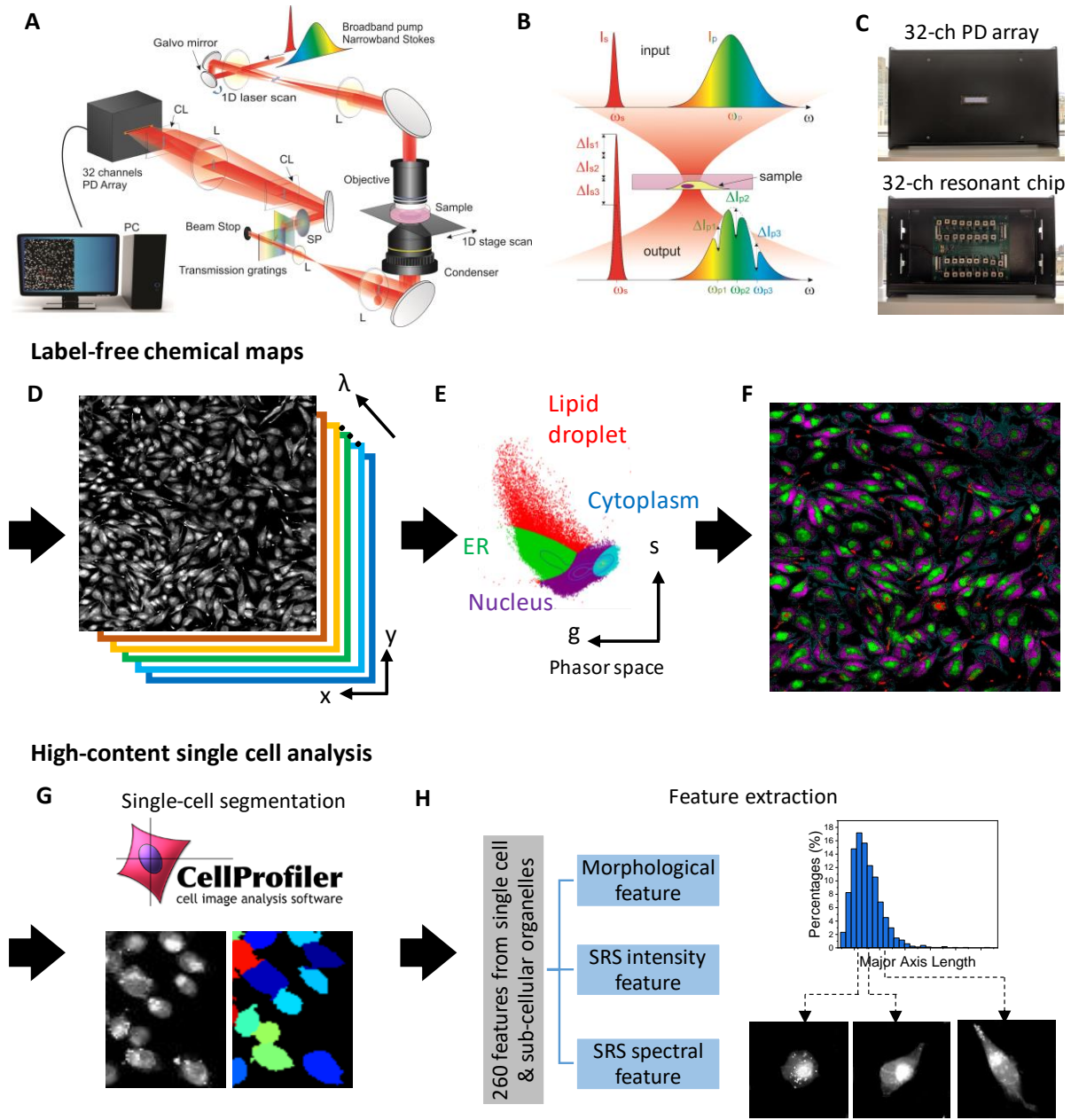

**Figure 1. Workflow of the single-cell analysis by the multiplex SRS imaging cytometry.**
(A-C), *In situ* spectroscopic imaging of a large number of cells by multiplex SRS. (A) A hybrid scanning was implemented by scanning a galvo mirror while moving the motorized stage. Multiple Raman modes are parallelly detected by a lab-built 32-channel lock-in free resonant photodiode array detector. (B) Multiple Raman shifts are excited by a broadband pump beam and a narrowband Stokes beam. (C) A photograph of our lab-built 32-channel lock-in free resonant photodiode (PD) array detector. Upper panel: the detector front side, showing a 32-channel PD array. Lower panel: the detector backside, showing 32-channel resonant circuit chips for lock-in free detection. (D-F) Label-free chemical mapping. (D) x-y-λ, a three-dimensional dataset generated by the multiplex SRS imaging cytometer. (E) The SRS spectrum from each image pixel is projected onto a 2-D phasor domain, followed by an unsupervised clustering algorithm to separate ER, nuclei, cytosol, and LDs. (F) A chemical map is generated by remapping the clustered results in the phasor domain back to the SRS image.

(G and H) High-content single-cell analysis. (G) Single-cell segmentation by CellProfiler. (H) Left panel: a total of 260 features in each cell are extracted, which can be classified into morphological features, SRS intensity features, and SRS spectral features. Right: statistical analysis of each feature demonstrates cellular heterogeneity.

**Multiplex SRS Imaging Cytometry Identifies Lipid-accumulated Protrusion as a Metabolic Signature for Cancer Cells under Stressed Conditions.** We applied our system to study a fundamental yet elusive question: how human cancer cells reprogram their metabolism to cope with a stressed microenvironment. We imaged six groups of human pancreatic cancer cells under a variety of stress conditions, including a chemotherapy stress model: (1) MIA PaCa-2 pancreatic cancer cell treated with gemcitabine drug; (2) MIA PaCa-2 and MIA PaCa-2 derived gemcitabine-resistant cell line G3K; (3) G3K cells treated with gemcitabine drug; and a starvation stress model: MIA PaCa-2 cells in glucose and serum-free medium for (4) 6 hours; (5) 12 hours; and (6) 24 hours. 260 features from three categories, including morphological features, SRS intensity features, SRS spectral features are extracted (**Figure S4**).

**Figure 2A** shows the extent of change of cellular features of MIA PaCa-2 or G3K cells under different conditions. From these feature heatmaps, we observed some common trends of morphological or metabolic change shared among gemcitabine-resistant G3K cells, gemcitabine or starvation treated cells. For example, gemcitabine treated, gemcitabine-resistant, and starved cells all tend to have increased major axis length (blue arrow), increase the LDs counts within ER (gray arrow), and increased mean distances of LDs out of ER to the center of the cell (green arrow). To better illustrate the distribution of these features in cell populations under different conditions, we elaborate individual features in histograms. **Figure 2B** demonstrates the distributions of major axis length (blue array in **Figure 2A**) of MIA PaCa-2 and G3K cells in different conditions. We found that gemcitabine treatment tends to broaden the distributions of both cell lines, with the maximum distribution shifting to a larger value; starvation of 12 hours or longer also broadens the distribution, but with no obvious shifting of central distribution. These results indicate that under stress condition, a subpopulation of cancer cells tend to have spindle-like morphology, which is likely due to the formation of protrusion structures. Another example is the stress-altered distribution of LDs in cells. **Figure 2C** displays the mean distance of LDs out of ER to the center of the cell (green array in **Figure 2A**), from which we find right shifts of the distribution in gemcitabine treated or resistant cells and in cells under starvation. Such a change indicates a spatial redistribution of LDs toward the distal location of cancer cells in stress conditions.

In order to better understand the relationships among the extracted 260 features, we generated a correlation matrix by calculating correlation coefficients from all features of the gemcitabine-resistant G3K cells (**Figure 2D**). From this table, we find some strong correlations between different morphological or metabolic features, most of which are features representing similar information. For example, we found that the protrusion features, such as major axis length, solidity, form factor, extent, compactness, and eccentricity strongly correlated with each other, presenting the morphological changes in multiple dimensions (**Figure S5**). Surprisingly, we also found that the protrusion features correlate with the LD features, such as the LDs location feature (**Figure 2E**). By plotting the major axis length of the cells to the average distance of LDs to the cell center, we found a strong positive correlation between them with a coefficient of 0.81. This suggests that in stress environments, the cells not only form protrusions but also have enriched LDs in the protrusions. This finding prompts us to perform a closer examination of the cells under SRS imaging. Consistent with our high-throughput analysis, we observed increased number of lipid-enriched protrusion structures in cells under a variety of stress conditions (**Figures 2F and 2G**), including gemcitabine treatment and starvation. Further experiments demonstrate that the number of LD-rich protrusions significantly increased in cells grown in either glucose-free or serum-free medium, and increased even further when both glucose and serum were deprived (**Figure S6A**). Quantitative analysis confirms a statistically significant increase of the percent of MIA PaCa-2 cells with LD-rich protrusions cultured in nutrition-deprived medium (**Figure S6B**). Moreover, this phenomenon not only occurs in pancreatic cancer cells but also occurs in cells of other types of cancer, including A549 lung cancer cells (**Figure S6C**) and MDA-MB-231 breast cancer cells (**Figure S6D**), suggesting LD-protrusions as a general response for the cancer cells under stress. We also confirmed that the formation of LD-protrusions is not dependent on cell density (**Figure S7**). It is worth noting that time-lapse SRS imaging reveals LDs transport from the perinuclear area of cells to the protrusions (**Figure S8**), indicating LD accumulation in protrusions is an active response of cancer cells under stress.

To summarize, by combining SRS imaging cytometry and high-throughput analysis of cellular features we identified both morphological and metabolic changes of cancer cells in stress conditions. Particularly, we found the formation of lipid-rich protrusion structure is a response of cancer cells to stress environment and a potential marker for stress-resistant cancer cells. The following sections are devoted to further analysis of this metabolic marker and its potential role in cancer cells.

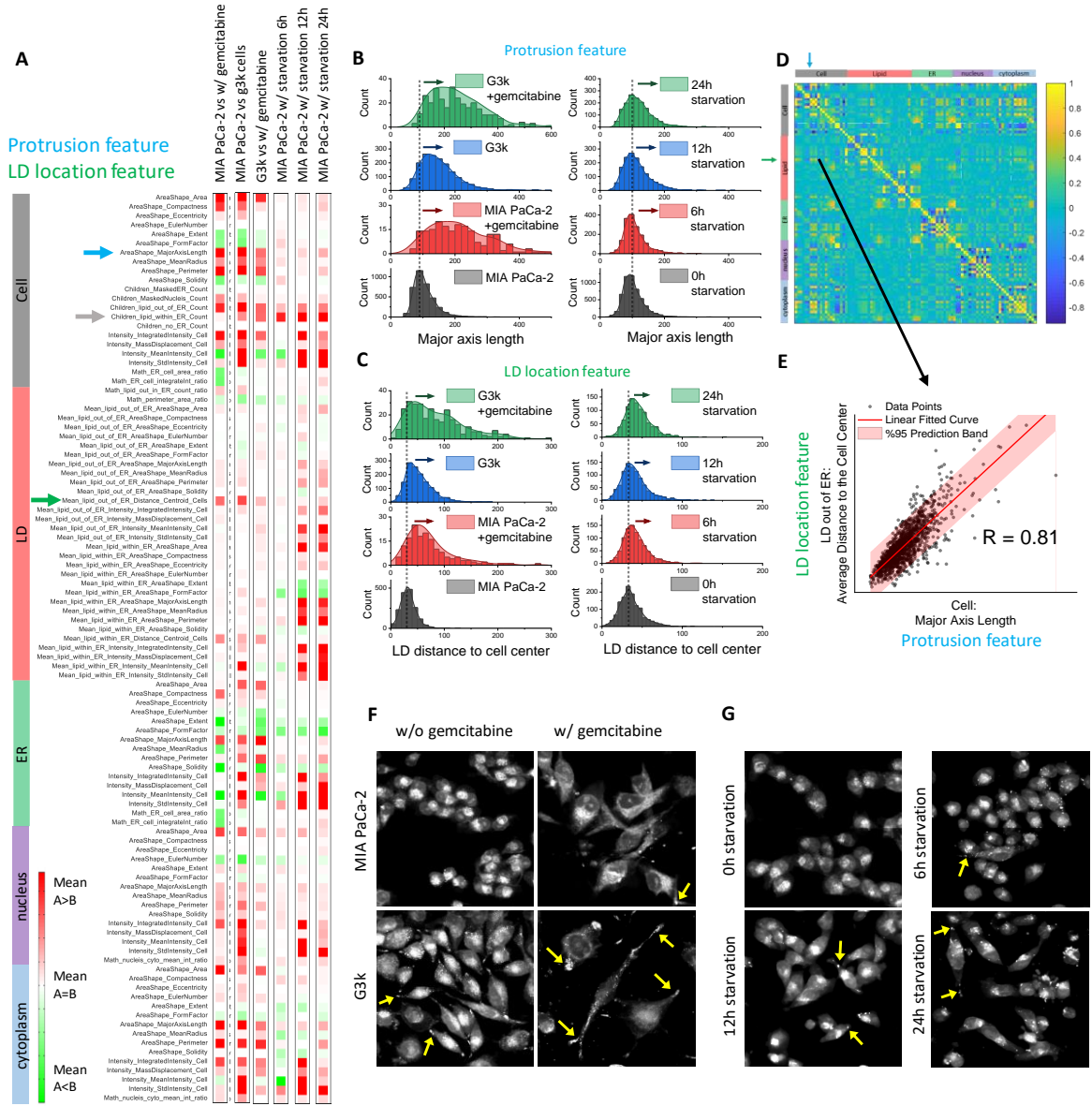

**Figure 2. Multiplex SRS imaging cytometry identifies unique metabolic signatures under stressed conditions.**

(A) Feature arrays of cells under chemotherapy stress model (column one to three) and starvation stress model (column four to six). Red or green color indicates a mean value higher or lower than the control group, respectively. From column one to six: MIA PaCa-2 cells treated with gemcitabine (n=642) compared with MIA PaCa-2 cells (control, n=1150); G3K cells (n=1637) compared with MIA PaCa-2 cells (control, n=1150); G3K cells treated with gemcitabine (n=313) compared with G3K cells (control, n=1637); MIA PaCa-2 cells starved for 6 hours (n=1698) compared with MIA PaCa-2 cell without starvation (control, n=5259); MIA PaCa-2 cells starved for 12 hours (n=1515) compared with MIA PaCa-2 cell without starvation (control, n=5259); MIA PaCa-2 cells starved for 24 hours (n=1547) compared with MIA PaCa-2 cell without starvation (control, n=5259).

(B) Histograms of the cell 'major axis length' for both the chemotherapy stress model (left panels) and the starvation stress model (right panels).

(C) Histograms of the 'distance of lipid droplets to the cell center', for both the chemotherapy stress model (left panel) and the starvation stress model (right panel).

(D) The correlation matrix of all the 260 features in gemcitabine-resistant G3K cells (n=1637). Yellow color indicates positive correlation coefficients with the highest value of 1. Blue color indicates negative correlation coefficients with the lowest value of -1.

(E) The 'major axis length' (a protrusion feature) is positively correlated with 'distance of lipid droplets to the cell center' (a lipid droplet feature), with a correlation coefficient of 0.81.

(F and G) Representative SRS images of cells from the (F), the chemotherapy stress model and (G), the starvation stress model. Yellow arrows indicate the lipid-accumulated protrusion structures.

**Multiplex SRS Imaging Cytometry Reveals a Location-dependent Heterogeneous Composition of LDs within a Single Cell.** Besides providing 260 morphology or intensity-based feature analyses, an irreplaceable advantage of our SRS imaging cytometry is to provide spectrum-based feature analysis, such as to distinguish heterogeneity of chemical compositions in LDs in a single cell. Location and content of LDs are dynamically regulated through processes involving synthesis on ER and degradation by lipophagy or lipases in the cytoplasm(Walther and Farese, 2012). High-speed SRS imaging cytometry allowed us to acquire a hyperspectral image of 339 G3K cells, possessing 7,193 lipid droplets, in 24 seconds. Among these LDs, we analyzed the compositional difference between LDs associated with ER (LDs within ER) and free LDs (LDs out of ER). The LDs out of ER showed different SRS spectrum compared to the LDs within ER, which can be separated in the spectral-phasor domain (**Figure 3A**). The spectral comparison highlighted a major difference at 2870 cm$^{-1}$ (**Figure 3B**). A cellular LD is mainly composed of triglycerides and cholesteryl esters. As triglyceride and cholesteryl ester have a major spectral difference at 2870 cm$^{-1}$ in the CH region (**Figure 3C**), we suspect that the spectral variation of LDs within ER and LDs out of ER at 2870 cm$^{-1}$ is due to the relative ratios between triglyceride and cholesteryl ester. To confirm such an LD compositional difference in a statistical manner, we performed SRS imaging of G3K cells (**Figure 3D**) and generated an LD-composition map of triglyceride/cholesteryl ester by 2930 cm$^{-1}$ to 2870 cm$^{-1}$ SRS ratio image (**Figure 3E**). We found that LDs out of ER tend to have less triglyceride compositions but more cholesteryl ester contents (**Figures 3F** and **3G**). Statistical analysis of all LDs in the image also showed a positive relationship between lower triglyceride content (higher cholesteryl ester content) with longer distance of LDs to the cell center (**Figure 3H**). In addition, the mean intensity of LDs is weaker with lower triglyceride content (**Figure 3I**) or farther away from the cell center (**Figure 3J**). These chemical content characteristics were also confirmed by the conventional spontaneous Raman spectroscopy measurement by analyzing the cholesterol signature peak around 704 cm$^{-1}$ (**Figures 3K** and **3L**). A quantitative analysis of spontaneous Raman signal at 704 cm$^{-1}$ is shown in **Figure S9**.

Collectively, the results indicate that even within a single cell, the contents of LDs are spatially different. The relative higher triglyceride/cholesteryl ester ratio in the LDs within ER suggests the deposition of triglyceride during LD synthesis, while the relatively lower ratio of LDs out of ER is possibly due to the degradation of triglyceride in the cytosol, especially in the protrusions. Having shown that LDs are translocating from the perinuclear part of the cell to the protrusions (**Figure S7)**, our results suggest LDs are under active utilization to facilitate protrusion growth.

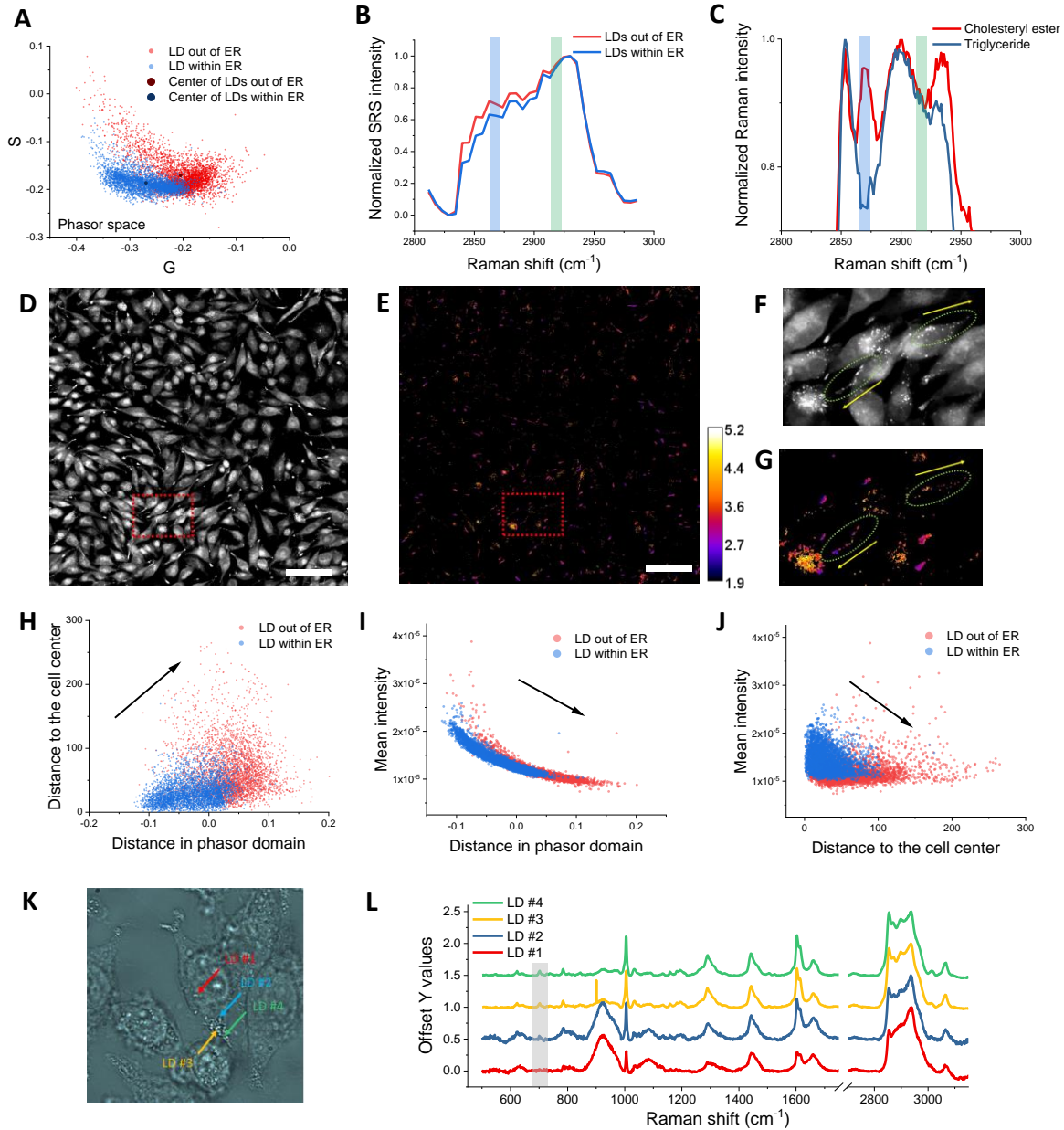

**Figure 3. Multiplex SRS imaging cytometry reveals the heterogeneity of LDs in single cells.**
(A) A scatter plot of LDs outside the ER (red) and inside the ER (blue) boundaries in the phasor space.
(B) Averaged SRS spectra of LDs outside the ER (red) and inside the ER (blue) boundaries in the C-H region.
(C) Spontaneous Raman spectra of cholesteryl ester (red) and triglyceride (blue) in the C-H region.
(D) An SRS image of G3K cells.
(E) The molecular composition map of LDs in G3K cells by 2930 $cm^{-1}$/2870 $cm^{-1}$ SRS ratio image. Lower values indicate more cholesteryl ester contents, and higher values indicate more triglyceride contents. Scale bar: 100 μm.
(F and G) A zoom-in SRS image (F), and a zoom-in composition map (G) of the dashed rectangular region in the image panels (D) and (E), respectively. Yellow arrows indicate the directions of LDs away from the cell center.
(H-J) 2D scatter plots of two features from LDs outside the ER (red) and LDs inside the ER (blue) boundaries. The 'distance in phasor domain' reflects 'lipid compositions' of the LDs. The higher value of 'distance in phasor domain' indicates more cholesteryl ester contents. The lower value of 'distance in phasor domain' indicates more triglyceride contents. The features are (H), 'lipid composition' vs 'the distance of LD to the cell center', (I), 'lipid composition' vs 'LD mean intensity', (J), 'distance of LD to the cell center' vs 'LD mean intensity'.
(K) A transmission image of G3K cells. Arrows indicate LDs analyzed by Raman spectroscopy.
(L) Spontaneous Raman spectra of selected LDs in the image (K). The gray region highlights the cholesteryl ester signature peak around 704 $cm^{-1}$.

**LDs in Protrusions are used for Local Energy Production.** Next, we aim to understand how LDs in protrusions are being degraded and what is the function of these LDs. LD degradation happens through lipophagy or lipolysis[15,16]. In lipophagy, autophagosomes engulf the LDs and fuse with lysosomes to breakdown the LDs(Thiam et al., 2013). In lipolysis, lipases including adipose triglyceride lipase (ATGL), hormone-sensitive lipase, and monoglyceride lipase are involved to degrade triglyceride sequentially to release free fatty acids(Thiam et al., 2013).

We first examined the involvement of autophagosomes for LD degradation at the cell protrusions. To monitor autophagy, we labeled autophagosomes with monodansylcadaverine, a fluorescent dye that can be excited by the 800 nm femtosecond pulses through the two-photon absorption. Then, we used duel-modality imaging, including the SRS microscopy and the two-photon excitation fluorescence (TPEF) microscopy, to simultaneously image the LDs and the autophagosomes, respectively. From the images, we found the presence of both LDs and autophagosomes in proximity in the LD-rich protrusions in MIA PaCa-2 (**Figure 4A**), G3K (**Figure 4A**), and A549 (**Figure S10A**) cells.

Excess free fatty acid released from LD degradation is toxicity if not properly used. They can be used for a variety of biological processes including energy production, membrane synthesis, or converted back to triglycerides. However, the enzymes related to membrane synthesis or triglyceride synthesis are majorly localized on the ER. Therefore, we hypothesize that the free fatty acids released at the cell protrusions would be used for energy production through mitochondria β-oxidation. We then studied the interaction of LDs in the cell protrusions with mitochondria. We labeled mitochondria with MitoTracker™ and imaged them by TPEF microscopy in parallel with the SRS imaging of the LDs. The duel-modality imaging reveals the co-presence of both organelles in proximity in the LD-rich protrusions in both MIA PaCa-2 (**Figure 4B**), G3K (**Figure 4B**), and A549 cells (**Figure S10B**). The evidence supports the local synthesis of ATP using LDs as an energy source.

The highly correlated the LD accumulation/degradation and cell protrusions implies a critical role of LD function for the cell protrusions. Since we have shown that LDs are likely to be used as an energy source, it is reasonable to hypothesize that LD supported local ATP production is a requisite for the formation of protrusions, and the inhibition of LD degradation would hamper the development of protrusions. To test this hypothesis, we treated cells with three inhibitors including chloroquine sulfate as an autophagy inhibitor, atglistatin as an ATGL inhibitor, and etomoxir as a β-oxidation (functions on carnitine palmitoyltransferase I, or CPT1) inhibitor in three protrusion induced models including starvation stress model, drug-resistant model, and chemotherapy stress model.

In the starvation model, we imaged and counted the total number of cells and the number of cells with protrusions treated with three inhibitors respectively at various concentrations in both normal and 48 hours starvation conditions. Quantitative analysis of percent of cells with protrusions show that chloroquine sulfate (**Figure 4C**), atglistatin (**Figure 4D**), and etomoxir (**Figure 4E**) significantly reduced the formation of LD-rich protrusions in cells under starvation condition, but not under normal condition, for both MIA PaCa-2 (**Figures 4C-4E**) and A549 cells (**Figures S10C- S10E**). These results support that LD degradation and association with mitochondria are critical for the formation of LD-rich protrusions under starvation when other energy sources are not present.

In the drug-resistant model, the protrusion structure is observed in G3K cells, a gemcitabine-resistant cell line (**Figure 2B**). We imaged G3K cells and G3K cells treated with chloroquine sulfate, atglistatin, and etomoxir. We found that each inhibitor effectively suppressed the formation of lipid-facilitated protrusions. The morphological feature of extent can be used to describe the cell protrusions. From the statistical analysis of the extent value of G3K cells, we found that although the cells show heterogeneous response to the inhibitors, the treatment tends to shift the extent distribution to larger values, indicating the decrease of the protrusions (**Figure 4F**). Other protrusion related features such as higher solidity value and lower compactness value show similar trends. (**Figures S11A- S11B**). In addition, we observed a decrease of LDs average distance to the cell center through statistical analysis. For the G3K cells, 36% cells showed a distance of LD to the cell center above average. This percentage dropped to 20%, 3.2%, and 23% after treated with chloroquine sulfate, atglistatin, and etomoxir, respectively (**Figure 4G**). These results indicate that LD degradation and fatty acid oxidation are critical for the formation of LD-rich protrusions in the chemotherapy-drug-resistant cells.

In the chemotherapy stress model, we performed the same inhibitor treatment described previously and obtained similar results (**Figures 4H, 4I, S11C, and S11D**). We found that each inhibitor effectively suppressed the formation of LD-facilitated protrusions in gemcitabine treated G3K cells. Collectively, our evidence shows that LDs are associated with energy production at the protrusions. Regulating the LD degradation and fatty acid oxidation processes impact the formation of protrusions in cancer cells in stress conditions. An illustration of the role of LDs at the protrusions is shown in **Figure 4J**.

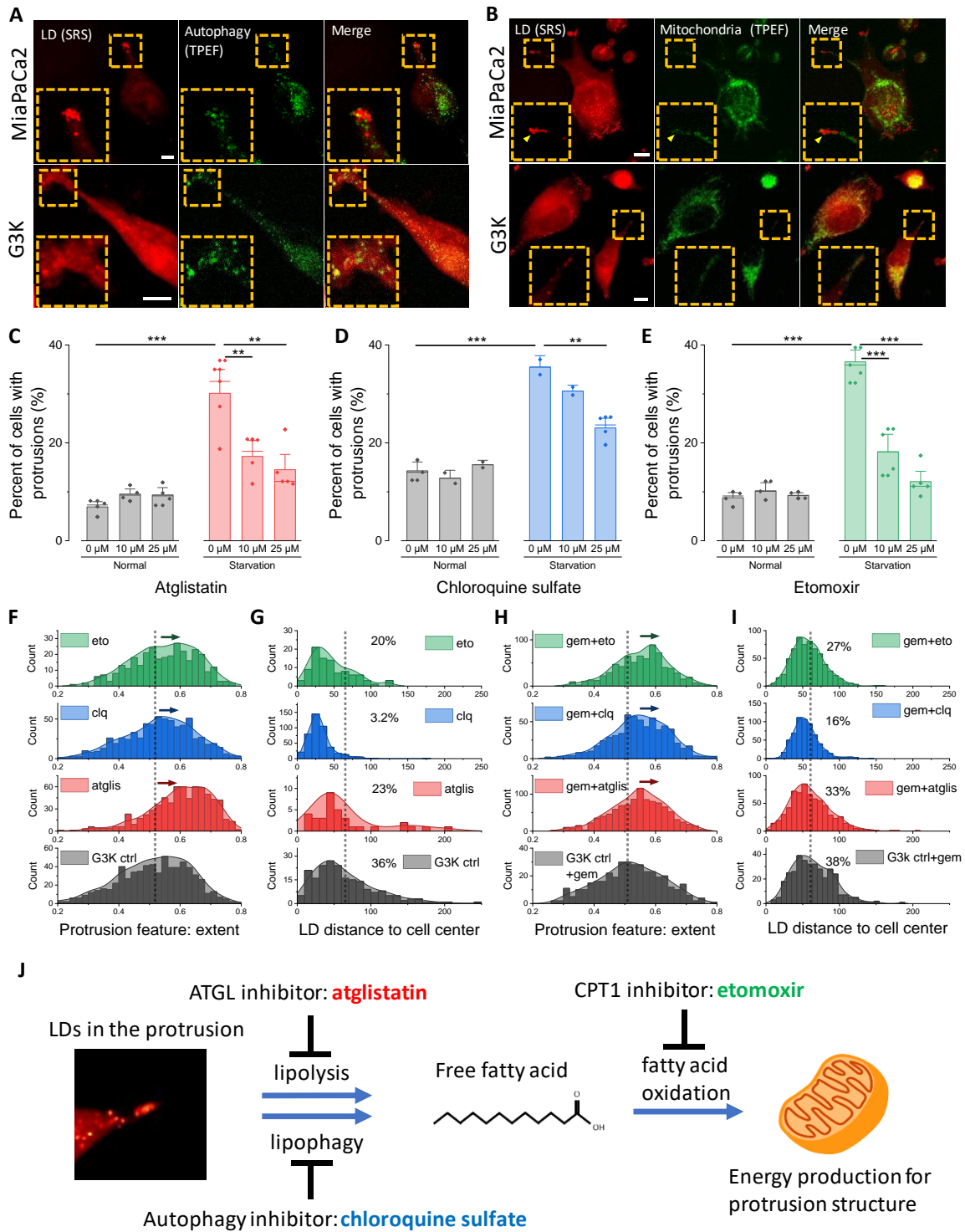

**Figure 4. LDs are degraded by autophagy and lipase and used in mitochondria for energy.**
(A) SRS (left panel), TPEF (middle panel), and the composition (right panel) images from a MIA PaCa-2 cell (upper panel) and a G3K cell (lower panel). TPEF imaging detects the autophagosomes labeled by monodansylcadaverine. Zoom-in areas from the dashed yellow square are shown in the lower-left corner of each panel.
(B) SRS (left panel), TPEF (middle panel), and the composition (right panel) images from a MIA PaCa-2 cell (upper panel) and a G3K cell (lower panel). TPEF imaging detects the mitochondria labeled by MitoTracker™. Zoom-in areas from the dashed yellow square are shown in the lower-left corner of each panel.
(C) Percentage of cells with LD-rich protrusions after treatment by atglistation in normal and starvation conditions (n=5).

(D) Percentage of cells with LD-rich protrusions after treatment by chloroquine sulfate in normal and starvation conditions (n=5).
(E) Percentage of cells with LD-rich protrusions after treatment by etomoxir in normal and starvation conditions (n=5).
(F and G) Histograms of one of the protrusion features 'extent' for (F) G3K cells and (G) gemcitabine treated G3K cells, treated with atglistatin (atglis), chloroquine sulfate (clq), and etomoxir (eto). The higher 'extent' value indicates cells with less protrusion formation.
(H and I) Histograms of the 'distance of LDs out of ER to the cell center' feature for (I) G3K cells and (J) gemcitabine treated G3K cells, treated with atglistatin, chloroquine sulfate, and etomoxir.
(J) Our hypothesis of LDs degradation at the protrusion by autophagy and lipase, and the utilization of free fatty acids for energy production via β-oxidation in mitochondria. The scale bars are 10 μm. ** $p < 0.01$, *** $p < 0.001$.

**Blockage of Lipid Degradation or β-oxidation Impairs the Survival of Stress-Resistant Cancer Cells.** The observation that the residual cancer cells under stress have longer protrusions implies an important role of protrusion formation for the cells to survive in a stressed environment. We tested whether targeting the altered lipid metabolism can eliminate stress-resistant cancer cells. We measured the viability of MIA PaCa-2 cancer cells under starvation and G3K cells treated by the three previously mentioned inhibitors, which have been shown to reduce protrusion formation. For the starvation-resistant cancer cells, we first tested the cell survival under starvation with or without inhibitor treatment and found that 50 μM autophagy inhibitor chloroquine sulfate, 50 μM lipase inhibitor atglistatin, or 50 μM β-oxidation inhibitor etomoxir induce a much faster cell death compared to vehicle treatments (**Figure 5A**). Then, we examined whether the starved cancer cells are more susceptible than the non-starved ones in response to the inhibitor treatments. We found that the starved cells are more vulnerable to treatment of chloroquine sulfate (**Figure 5C**), atglistatin (**Figure 5D**), or etomoxir for 24 hours (**Figure 5E**). Similarly, the treatment of inhibitors substantially diminished the survival of A549 cells under starvation condition (**Figure S12A**) but not normal condition (**Figures S12B, S12C, and S12D**).

We also tested whether drug-resistant cancer cells are more susceptible to synergistic treatment. The results show that synergistic treatment of gemcitabine drug with 50 μM autophagy inhibitor chloroquine sulfate, 50 μM lipase inhibitor atglistatin, or 50 μM β-oxidation inhibitor etomoxir induce a much faster cell death compared to vehicle treatments (**Figure 5B**). In addition, we found that the gemcitabine-resistant cancer cells are more vulnerable to chloroquine sulfate at 72 hours (**Figure 5F**), atglistatin at 72 hours (**Figure 5G**), and etomoxir at 72 hours (**Figure 5H**).

These results collectively demonstrate that stress-resistant cancer cells rely on LD degradation and fatty acid oxidation processes for survival. Since the inhibitors also reduce the formation of protrusions (**Figure 4**), our results indicate the important role of LD-rich protrusions in the survival mechanism of stress-resistant cancer cells.

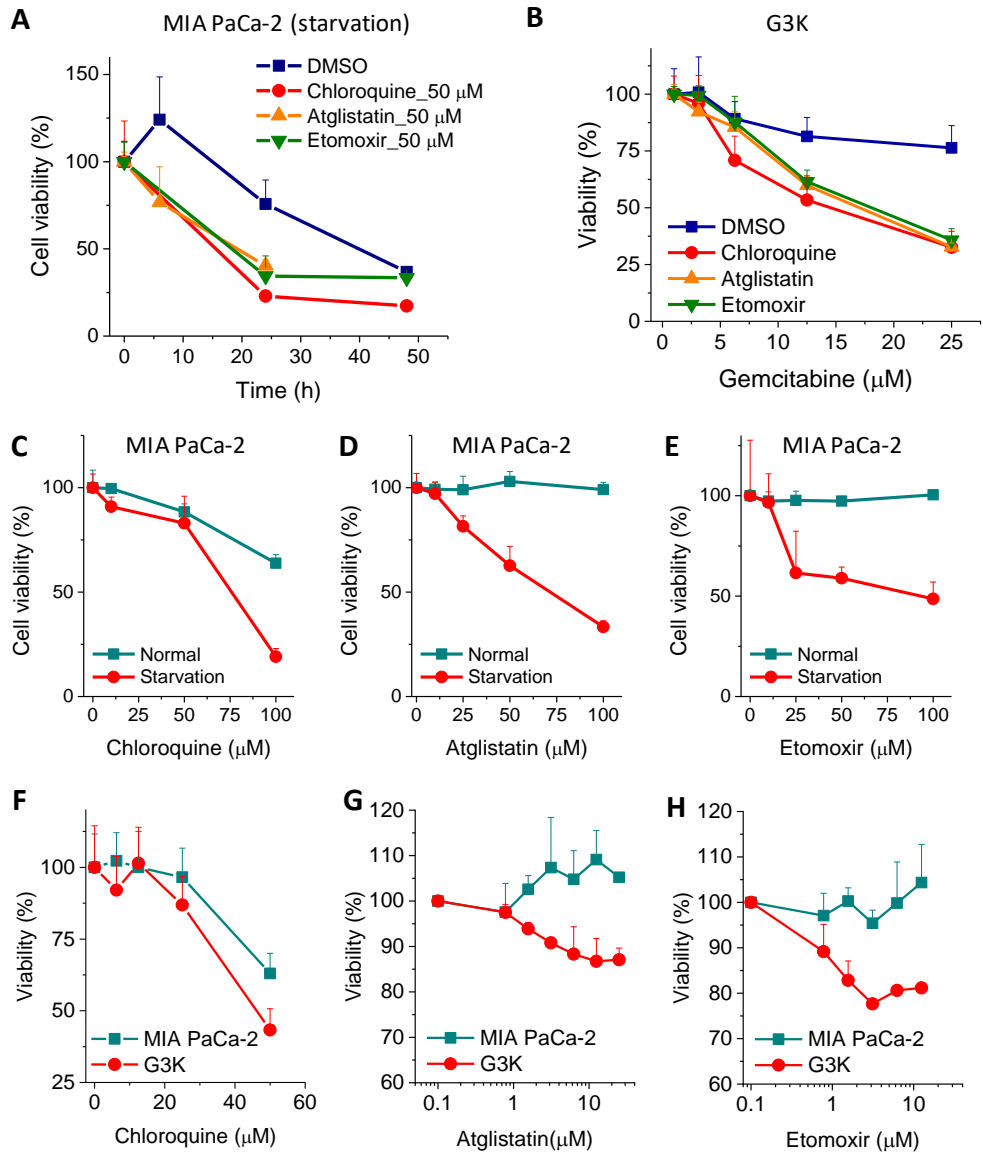

**Figure 5. Blockage of lipid metabolism suppressed cell survival under stress.**
(A) Time-dependent MIA PaCa-2 cell viability with and without chloroquine sulfate treatment, atglistatin treatment, and etomoxir treatment under starvation condition.
(B) Gemcitabine concentration-dependent G3K cell viability with and without chloroquine sulfate treatment, atglistatin treatment, and etomoxir treatment for 72 hours.
(C-E) Concentration-dependent MIA PaCa-2 cell viability for cells treated by (C) chloroquine sulfate, (D) atglistatin, and (E) etomoxir under normal and starved conditions for 24 hours.
(F-H) Concentration-dependent viability for MIA PaCa-2 cells and G3K cells, treated by (F) chloroquine sulfate, (G) atglistatin, and (H) etomoxir, for 72 hours. n=3.

**LD-rich protrusions enhance glucose uptake.** The formation and growth of LD-rich protrusions can significantly alter the shape and size of the cells. As mentioned earlier, the elongated protrusions rendered cells more polar-shaped (**Figures 2B-D, and S4**). This phenotypic change is likely to extend the territory of the cell and potentially enhance the uptake of nutrition from the medium, consequently promoting cell survival. To test this supposition, we applied starvation model and measured glucose uptake in the starved cells for 24 hours to stimulate the growth of LD-rich protrusions. Then, we fed the cells with a fluorescent glucose analog, 2-(n-(7-nitrobenz-2-oxa-1,3-diazol-4-yl) amino)-2-deoxyglucose (2-NBDG) for 1 hour, and imaged 2-NBDG uptake by TPEF microscopy in the cells. In both MIA PaCa2 (**Figures 6A, and B**) and A549 cells (**Figure S13**), starved cells with LD-rich protrusions showed much higher 2-NBDG uptake than cells without LD-rich protrusions or non-starved cells. Furthermore, we performed immunostaining and imaging of glucose transporter GLUT1 in normal and starved MIA PaCa2 cells. Compared to normal cells, starved cells have a significantly higher expression level of GLUT1 (**Figure 6C**). Moreover, GLUT1 is highly expressed at the membranes of the protrusions in starved cells. In contrast, GLUT1 expression mostly localizes to cell body area in normal cells (**Figure 6C**). We also tested the supposition in the chemotherapy drug model and measured the glucose uptake of cells treated with gemcitabine to stimulate the growth of LD-rich protrusions. In both MIA PaCa2 (**Figure 6D**) and G3K cells (**Figure 6D**), gemcitabine-treated cells showed much higher 2-NBDG uptake than cells without gemcitabine treatment. In addition, G3K cells, which are observed with more LD-rich protrusions, obtain much higher 2-NBDG uptake than MIA PaCa2 cells. The evidence suggests that lipid-rich protrusions as a metabolic reprogramming signature may play an important role for cells to acquire glucose more efficiently. Although our preliminary data showed that the accumulation of LDs in protrusions is associated with the AMPK signaling (**Figure S14**), the molecular mechanisms need to be further studied.

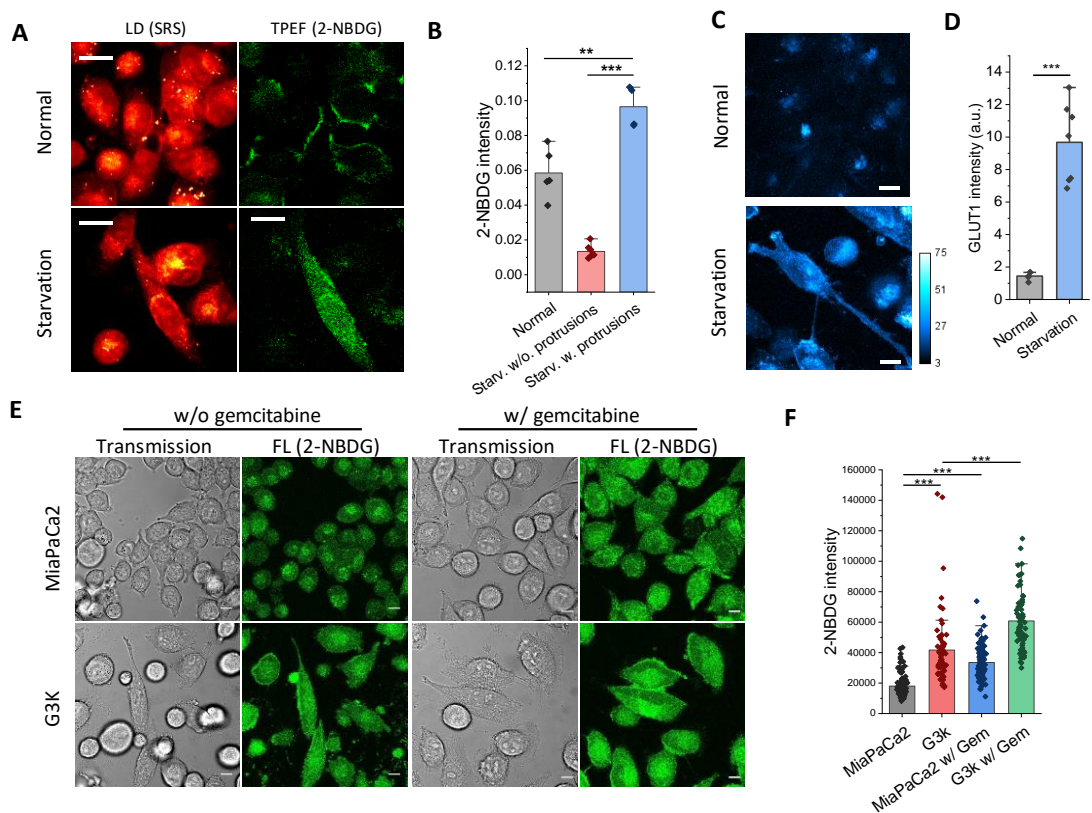

**Figure 6. Protrusions increase cellular uptake of glucose.**
(A) SRS (left panels), and TPEF (right panels) images from MIA PaCa-2 cells incubated with 2-NBDG for 1 hour. The TPEF signal was from the 2-NBDG accumulated in the cells. The upper panels are the control group, while the lower panels are collected under the starvation condition.
(B) Quantitation of the 2-NBDG intensity for MIA PaCa-2 cells in normal (n=5) and starvation conditions (with (n=6) and without protrusions (n=7)).
(C) Representative TPEF images of MIA PaCa-2 cells immunostained with an antibody against GLUT1. Scale bar: 10 μm.

(D) Quantitation of fluorescent intensity from labeled GLUT1 in panel (C) (n = 6).
(E) Transmission phase contrast (left panels) and TPEF (right panels) images of MIA PaCa-2 and G3K cells incubated with 2-NBDG for 2 hours.
(F) Quantitation of the 2-NBDG intensity for MIA PaCa-2 and G3K cells without (MIA PaCa-2: n=105; G3K: n=67) and with gemcitabine treatment (MIA PaCa-2: n=103; G3K: n=68).

**Discussion**

Cancer cells are commonly exposed to stress environments. For example, due to the insufficient blood distribution, cancer cells at the center of a tumor often suffer from low oxygen and nutrition supply; chemotherapy drugs create significant challenges to cancer cells. However, a small portion of cancer cells manages to survive in the stress environments by reprogramming their metabolism, which may lead to incomplete tumor remission and eventually tumor relapse. Therefore, understanding the metabolic strategies utilized by cancer cells under stress conditions is critical to gain insights for developing more effective and targeted cancer therapy.

To address this challenge, a new tool which can perform high-speed metabolic imaging and high-throughput analysis of cell metabolism is essential. We first achieve technical innovation by developing an *in situ* chemical imaging cytometer enabled by high-speed multiplex SRS imaging. This approach allows us to measure a large number of cells with immense information in cellular or subcellular contents. After separating different subcellular compartments, including ER, nucleus, LDs and cytoplasm at high resolution, we were able to extract 260 morphological or metabolic features to understand and quantify cellular metabolic changes. Our method addresses challenges caused by cellular heterogeneity from a novel angle, including resolving cellular heterogeneity in spatial distribution and in composition (for example, triglyceride and cholesterol contents), which is beyond the scope of conventional fluorescence-based techniques. Taking advantage of our imaging cytometry, we discovered LD-enriched protrusions as a metabolic marker for cancer cells under stresses, highlighting the potential of targeting lipid metabolism for effective elimination of starvation-resistant or chemotherapy-resistant cancer cells.

We see this resource opens a variety of applications. Our multiplex SRS imaging cytometry and pipelines not only apply for analyzing cultured cells, but also apply for *in situ* chemical mapping of larger biological objects, i.e. embryos, pathological tissue samples, and even whole organisms such as drosophila, zebrafish, and etc. Enabling by the hybrid scanning capacity, our multiplex SRS imaging cytometry can map both the distributions of chemicals in individual cells throughout a bulk specimen, in addition to the morphological information. With such information, we can examine the histopathological status of tissues from patients with diseases, or discover new potential biomarkers for disease identification. Moreover, assisted by deep learning we can potentially identify different cell types in complex tissue environments or in circulating systems. It allows studies of cell-to-cell metaboic interactions in situ, which is not possible with current techniques.

Despite the power that multiplex SRS imaging cytometry has demonstrated in metabolic discovery, the capacity of SRS imaging cytometry can be further improved in the following aspects. One limitation of SRS imaging cytometry is sensitivity, which could be improved with the following methods. (1) Using a nonlinear spectral compression scheme (Liu et al., 2015), imaging sensitivity can be enhanced by improving the pulse shaping efficiency without sacrificing the spectral resolution. (2) Using pre-resonance SRS, electronic resonant SRS or plasmonic-enhanced method, the Raman cross-section of a molecule could be further enhanced for detection of molecules at sub-micromolar concentration (Shi et al., 2018; Wei et al., 2017; Zhang et al., 2014).

Segmentation of intracellular compartments, organelles, or even molecules could be improved with better chemical selectively, which could be another perspective for future development. (1) Using a laser system with a broader spectral width, the Raman spectral window can be widened (Figueroa et al., 2018; Hiramatsu et al., 2019). (2) Combining with Raman tags which have much higher Raman scattering cross-section than most inherent biomolecules, the multiplex SRS imaging cytometry would be able to measure more biomolecules such as amino acids, DNA/RNA molecules, glucose, and small-molecule drugs, etc (Zhao et al., 2017). (3) Shifting currently NIR laser excitation to visible range is expected to improve the spatial resolution down to 130 nm to visualize fine structure such as dendritic spine (Bi et al., 2018). (4) It can be further extended to high-parameter cytometry by combining multiplex SRS cytometry and super-multiplex (24-colour) Raman probes to overcome the color barrier limited by intrinsically broad fluorescence spectra (Wei et al., 2017).

In our study, we showed LDs in protrusions are majorly used to produce energy in mitochondria to support the protrusion growth. However, serving as an energy source may not be the only possible function of lipids. We also tested the possible fate of lipids in LDs for membrane synthesis by using BODIPY-C12, a metabolizable fluorescent fatty acid analog and precursor for phospholipids and neutral lipids (Kolahi et al., 2016). In a previous report, BODIPY-C12 was shown to transport from LDs to organelles under starvation in fibroblast cells (Rambold et al., 2015). Here, using a similar approach, our results do not show the presence of BODIPY-C12 on plasma membrane after 12 or 24 hours starvation (**Figure S15**), suggesting that the fatty acids released from protrusion LDs are not used for membrane synthesis. On the other hand, inhibition of lipid β-oxidation pathway significantly suppresses protrusion formation, supporting lipids in protrusion-LDs as an energy source. Transportation of LDs from cell body to protrusions is meaningful and necessary in the sense that co-localization of LDs, autophagosome/lipase, mitochondria can maximize the efficiency of ATP production to maintain a favorable ATP concentration to support the fast expansion of the protrusions. However, the same strategy may not be applicable for efficient membrane construction. Synthesis of phospholipids is an energy-consuming process, which occurs in the ER. As a complex network structure, ER is primarily located in the cell body. Synthesis of phospholipids in the ER and transport them to the protrusion over long distance is low-efficiency and energy-unfavorable. In contrast, a more efficient way for membrane growth under starvation is probably through degradation and fusion of organelle membranes, which is consistent with the previous report that ER and other organelles are degraded under starvation condition(Nakatogawa, 2016). Future studies will be performed to fully understand the process of membrane construction under stress condition.

Although it is known that cancer cells under starvation or chemotherapy treatment become more vulnerable to other drugs, the inhibitors of autophagy/lipase/β-oxidation not only suppressed cell viability, but also selectively reduced protrusion formation under stress, but not under normal condition. This unique correlation suggests these inhibitors act through a specific protrusion-associated mechanism, but not due to general chemical cytotoxicity. It is likely that the inhibitors will distribute through the entire cell instead of just localizing at the protrusions. However, after prolonged starvation (24 h or longer) or chemotherapy treatment, the majority of LDs have been transported from the cell body to the protrusions. It is, therefore, reasonable to consider that the inhibitors play their roles primarily at the protrusions.

Our current research explored the behaviors of cultured cancer cell-lines on a two-dimensional surface. The same methodology could be applied to three-dimensional cell culture or tissues for *in situ* analysis of metabolism at the single-cell level. A more delicate segmentation algorithm might be needed to segregate individual cells and distinguish different cell types in a complex tissue environment. In addition to starvation stress and chemotherapy stress, other metabolic stresses, such as hypoxia or oxidative stress, might also impose unique traits on metabolism of cancer cells. For example, hypoxia condition may impair the capability of cancer cells to consume lipids through β-oxidation in mitochondria. In this case, alternative energy sources might be employed through different metabolic pathways. Future studies will be conducted to depict these metabolic adaptions in specific niches.

## Acknowledgements

We thank Dr. Chien-Sheng Liao and Gregory Eakins for the help and the design of TAMP array. This work was funded by National Institutes of Health (NIH) grants GM118471 and CA223581 to J.-X.C.

## Author contributions

K.-C.H., J.L. and C.Z. designed the experiment. K.-C.H., J.L. and C.Z. performed SRS imaging. J.L. performed cell viability experiment. K.-C.H., J.L. and C.Z. analyzed the data. J.L. and T.Y prepared the cell lines and the treatments. K.-C.H., J.L. and C.Z. wrote the manuscript. J.-X.C. supervised the project. All authors discussed the results and contributed to the manuscript.

## Declaration of interests

The authors declare that they have no competing interests.

## Methods

### Label-free chemical imaging cytometry by multiplex SRS

The laser source is a femtosecond laser (InSight X3, Spectra-Physics, Santa Clara, CA, USA) with two synchronized outputs at a repetition rate of 80 MHz. A Stokes beam is fixed at 1045 nm with output power ~3.5W, and a pump beam is tunable ranged from 680 nm to 1300 nm with output power ~1.0W. The Stokes beam is modulated at 2.3 MHz by an acousto-optic modulator (522c, Isomet Corporation, USA). In order to realize multiplex SRS, we utilize a narrow-band Stoke and a broad-band pump to simultaneously excite multiple Raman shifts. The bandwidth of the Stokes beam is spectrally narrowed to 13.4 $cm^{-1}$ by a pulse shaper described previously(Zhang et al., 2017). A broad pump beam with ~120 fs pulse width corresponds to the SRS spectral window ~200 $cm^{-1}$. The two laser pulses are spatially overlapped by a dichroic mirror and temporally overlapped by a translation delay stage. The combined lasers are guided to microscope modified based on a commercial microscope frame (BX51, Olympus, Japan). A 4-f lens system conjugates the galvo scanning plane to the objective entrance and expands the beam width to fulfill the back aperture of the objective. A water objective (UPLSAPO 60XW, NA=1.2, Olympus, Japan) is used to focus the laser into the sample at a diffraction-limit spot. The pump and Stokes laser powers are 40 mW and 60 mW on the sample. The pixel dwell time is set as 5 μs to acquire a multiplex SRS spectrum. An oil condenser with high NA value is used to collect transmitted pump beam to avoid cross-phase modulation background and to ensure efficient signal collection. After that, the pump beam is dispersed by two diffraction gratings (1200 groove $mm^{−1}$), and parallelly collected by our lab-built lock-in-free 32-channel photodiode array detector(Zhang et al., 2017). In order to avoid spectrum distortion, the incident angle of the light toward the dispersion grating is static during the image scanning. Therefore, an *in situ* hybrid scanning schematic is implemented. The fast axis is scanned by a galvo mirror in the direction perpendicular to the dispersion plane; whereas the slow axis is scanned by a motorized stage (PH117, Prior Scientific) moving in the direction the same as the grating dispersion. This hybrid scanning scheme, combined with large-area mapping scanned by a motorized scanning stage, allows us to acquire an x-y-λ data cube (2184 × 2184 × 32 voxels) spatially sampled of 250 nm with the region of 590 μm × 590 μm × 200 $cm^{-1}$ in 24 seconds.

**Cell lines and chemicals**

Human cancer cell lines, including MIA PaCa-2, A549, and MDA-MB-231, were obtained from the American Type Culture Collection. RPMI and DMEM cell culture media, fetal bovine serum (FBS), phosphate-buffered saline (PBS), BODIPY™ FL C12 and MitoTracker™ Green FM were purchased from Thermo Fisher Scientific. Chemicals including chloroquine sulfate, etomoxir, atglistatin, 2-(n-(7-nitrobenz-2-oxa-1,3-diazol-4-yl)amino)-2-deoxyglucose (2-NBDG) and 5-Aminoimidazole-4-carboxamide ribonucleotide (AICAR) were purchased from Selleck Chemicals or Cayman Chemicals. Dimethyl sulfoxide (DMSO), formaldehyde, and monodansylcadaverine (MDC) were purchased from Sigma Aldrich.

For the control groups, MIA PaCa-2 cells were cultured in RPMI 1640 + 10% FBS + penicillin/streptomycin (p/s) medium, A549 and MBA-MD231 cells were cultured in DMEM + 10% FBS + p/s medium. G3K cells were cultured in the same media supplemented with 1.0 μM gemcitabine to maintain the resistance. To create the starvation condition, glucose and serum-free DMEM medium were used, unless otherwise indicated. To create the chemotherapy treatment condition, 10 μM gemcitabine was used to treat MIA PaCa-2 cells for 24 hours, and 50 μM gemcitabine was used to treat G3K for 24 hours. For maintenance, all cells were cultured at 37 ºC in a humidified incubator with 5% $CO_2$ supply. Gemcitabine was purchased from Selleckchem (Catalog No.S1714).

**Label-free chemical mapping of sub-cellular organelle**

Spectral phasor analysis is implemented on the denoised data cube to map the high-dimensional SRS spectrum of each pixel into a vector in a two-dimensional phasor space(Fu and Xie, 2014). The distance of two phasor coordinates is determined by their spectral similarity. With the fact that each sub-cellular organelle obtains similar Raman spectra, Gaussian mixture model (GMM) is applied to automatically cluster the vectors in the phasor space into four main groups, including nucleus, ER, LDs, and cytoplasm. The clustered points in the phasor domain are then mapped back to an area in the cell image to generate the chemical mapping of each sub-cellular organelle. Quantitative phasor analysis combined with unsupervised GMM were analyzed by MATLAB (MathWorks).

**Cell segmentation and feature extraction**

We used CellProfiler, a free open-source software, to segment clumped cell and extract morphological features(Carpenter et al., 2006). SRS summed intensity cell image and the four sub-cellular mappings are sent to CellProfiler as the input images, followed by the pipelines customized for the analysis. For the cell segmentation, we utilized nuclei and ER as a starting point (primary object), and then identify the border of cell (secondary object). This strategy allows us to separate cell from a clumped environment. The accuracy of segmentation results was manually checked. The feature extraction of each sub-cellular organelle and the cell were also realized by CellProfiler. Morphology-based feature including lipid location and protrusion features are quantitatively measured by the MeasureObjectSizeShape module. Intensity-based feature are quantitatively measured from the MeasureObjectIntensity, MeasureObjectIntensityDistribution module. A spectral-based feature is realized with averaged SRS spectrum from the mask of each sub-cellular organelle. Parameters retrieved from morphology, intensity, and spectrum measurements from the cell and four sub-cellular organelles were then merged together (yielding 260 variables) for the features to describe single cell. Histogram of the selected feature was plotted using Origin 2017 for display.

**Femtosecond pulse SRS and two-photon excitation fluorescence (TPEF) imaging**

Femtosecond pulse SRS images in **Figures 5, 7, and 14,** and colocalization experiment with TPEF imaging was performed by the setup similar to the previous published.(Zhang et al., 2011) Our lab-built SRS microscope was constructed on a modified upright microscope frame (BX51, Olympus Corporation). The SRS was excited by a femtosecond laser system (InSight DeepSee, Spectral Physics). The center frequencies of the pump and Stokes beams were tuned to 800 nm and 1040 nm, respectively, to excite the $CH_2$ stretching vibration at ~2850 $cm^{-1}$. This Raman transition is majorly from the lipid molecules in cellular LDs. The pump and Stokes beams were collinearly overlapped and combined, with the Stokes beam modulated at 2.3 MHz by an acousto-optic modulator (1205-C, Isomet Corporation). The power of the pump and Stokes beams at the sample was, respectively, 10 mW and 20 mW. The laser beams were scanned using a 2D laser scanning system (GVSM002, Thorlabs) to form images. The beams were focused using either a 40X(LUMPLFLN 40XW, NA=0.8, Olympus) or a 60X(UPlanApo/IR 60XW, NA = 1.2, Olympus Corporation) water immersion objective onto the sample. The typical pixel dwell time was 10 µs. A high NA oil condenser (NA=1.4) was used to effectively collect the transmitted beams. A short-pass filter (980SP, Chroma) was used to reject the Stokes beam from entering into the photodetector. The pump beam was detected by a large-area photodiode (S3994, Hamamatsu Photonics) with an alternate current (AC) and a direct current (DC) output. The AC signal was sent to a digital lock-in amplifier (HF2LI, Zurich Instrument) synchronized to the function generator at 2.3 MHz for SRS signal extraction and amplification.

The TPEF imaging was performed on the SRS laser scanning microscopy platform. For the fluorescent probes used in this work, the 800 nm beam was effective for the two-photon excitation. To achieve parallel two-channel imaging (SRS and TPEF), a dichroic mirror was used to separate the TPEF signal from the transmitted beam for detection by a photomultiplier tube (PMT) (H422-40, Hamamatsu Photonics). After the PMT, a pre-amplifier was used to amplify the signal before sending the signal to the data acquisition card (PCIe-6363, National Instruments).

**Spontaneous Raman spectroscopy**

Confocal Raman was performed by a commercial Raman microscope (LabRAM HR Evolution, Horiba) at room temperature. A 15 mW (after the objective), 532-nm diode laser was used to excite the sample through a 40× water immersion objective (Apo LWD, 1.15 N.A., Nikon). The total data acquisition time was 60 s using the LabSpec 6 software. For all the spontaneous Raman spectra, we subtracted the PBS solution background. Cholesteryl ester percentage in lipid droplets was linearly correlated with the 704 $cm^{-1}$ peak normalized with the 2,850 $cm^{-1}$ peak. Cholesteryl oleate and glyceryl trioleate were purchased from Sigma-Aldrich.

**Image processing and statistical analysis**

The SRS and TPEF images were obtained by a lab-developed software based on LabView. The majority of images displayed in the figures were 400 × 400 pixels. For the large-scale SRS images, each image pane was 400 × 400 pixels. Raw images in ASCII format were converted to 16-bit PNG files for display. ImageJ was used to process the images. For the SRS images, the color scheme 'Red Hot' was used for display, and only the contract and brightness were optimized. The TPEF images were converted to 'Green' color scheme for display.

**Cell viability assay**

Cell viability was measured by Thiazolyl Blue Tetrazolium Blue (MTT) colorimetric assay (Sigma Aldrich), or Cell Counting Kit-8 (Dojindo Molecular Technologies, Inc), or CellTiter-Glo Luminescent Cell Viability assay (Promega Corporation). Cells were seeded in 96-well plates and incubated overnight. Then starvation and/or inhibitor treatments were performed for the indicated time period. The results are shown in Means + SD. N = 3-6 each group.

**Treatment of cells by chemicals**

To image the autophagosomes together with LDs, we labeled autophagosomes with MDC. The cells were first seeded and incubated overnight in 5% $CO_2$ + 37°C environment. Then the cells were incubated with 50 μM MDC for 15 min. Before imaging, the cells were triple-washed with the culture media to remove the MDC residues. The fluorescence signal from the MDC was collected by the TPEF imaging system. Similarly, mitochondria tracker (CELL) was used to image the mitochondria by the TPEF imaging system. The final concentration of MitoTracker™ was 250 nM, and the incubation time was 30 min.

To treat cells with inhibitors, the inhibitors were first mixed with cell culture media at final concentrations. Then the original culture media were switched to the prepared media at specific time points. Before treatment, all cells were cultured at least overnight in the incubator to allow adhesion to the culture dish. Cells were treated with chloroquine sulfate, atglistatin, etomoxir, or AICAR at concentrations indicated in figure captions. After treatment, the cells were washed with PBS triple times before imaging.

For the nutrition uptake analysis, we treated the cells with 2-NBDG at 100 μM for 2 hours. The cells were then triple-washed by the culture media before imaging. Live cell imaging was performed to analyze the uptake of 2-NBDG. For the analysis of membrane lipid conversion (**Figure S15**), BODIPY-C12 treatment was performed at a concentration of 2 μM for 16 hours. Triple-wash using PBS was also performed before imaging. The cells were also imaged alive.

**Immunofluorescence staining of GLUT1**

Cancer cells cultured in 35 mm glass-bottom dishes under normal or starvation condition for 24 hours were fixed in 10% neutral buffered formalin solution for 30 minutes. Immunofluorescence staining was performed following the manufacturer's protocol. Rabbit polyclonal antibody against glucose transporter GLUT1 was purchased from Abcam (Cat #ab15309) and used with 1:100 dilution. Alexa Fluor 488 conjugated goat anti-rabbit IgG (H+L) (ThermoFisher, Cat #A-11034) was used as secondary antibody with 1:200 dilution. TPEF imaging was performed using 800 nm femtosecond laser as excitation and the signal was collected with PMT with a 520/40 filter.

**Denoise image pre-processing**

Denoise algorithm was implemented to remove the noise in the x-y-λ data cube via total variation minimization(Liao et al., 2015b). The parameters of $β_x$, $β_y$, and $β_λ$ were determined corresponding to the spatial and spectral resolution of the system. The code is also available from Matlab FileExchange: http://www.mathworks.com/matlabcentral/fileexchange/52478-the-spectral-total-variationdenoising-algorithm.

**Statistical analysis**

For the quantitative analyses performed in this paper, one-way student's t-test was used for comparisons between groups. Results were represented as means + standard deviation. Statistical significance was indicated as * $p < 0.05$, ** $p < 0.01$, and *** $p < 0.001$.

**Data and software availability**

The raw data obtained and the custom codes for image analysis and sort-decision making are available from the corresponding author upon request.

**Supplementary Information**

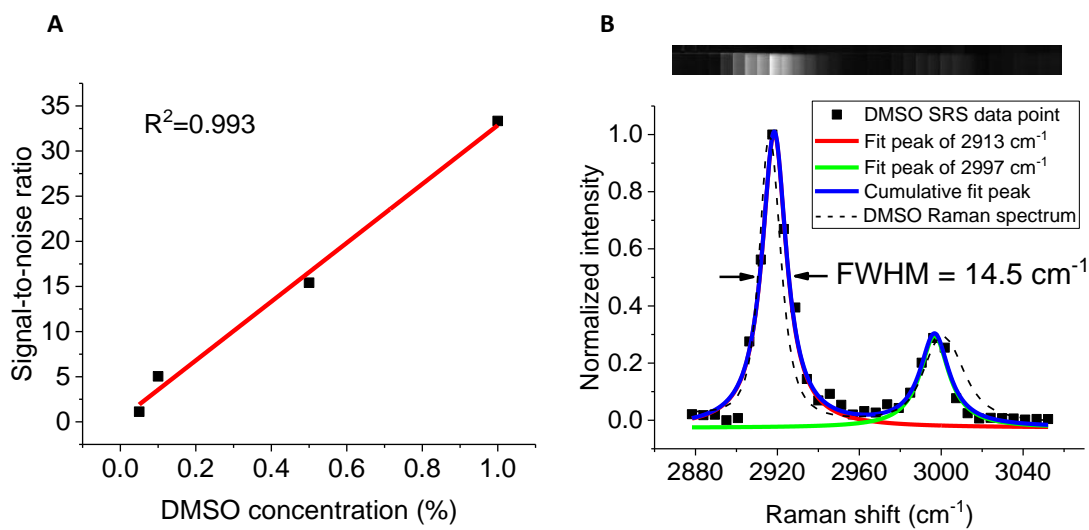

**Figure S1. System characterization of sensitivity and spectral resolution of multiplex SRS imaging cytometry.**

(A) DMSO sensitivity test. The detection limit is 0.1% DMSO concentration diluted in D2O solution.
(B) SRS (dot), SRS by Lorentzian fitting (solid line), and spontaneous Raman (dashed line) spectra of DMSO. SRS imaging of DMSO/air interface is shown above. The full width at half maximum (FWHM) of 2913 cm$^{-1}$ peak is measured as 14.5 cm$^{-1}$. After deconvolution of nature bandwidth of DMSO peak with 5.4 cm$^{-1}$, the spectral resolution of multiplex SRS imaging cytometry is calculated as 13.4 cm$^{-1}$.

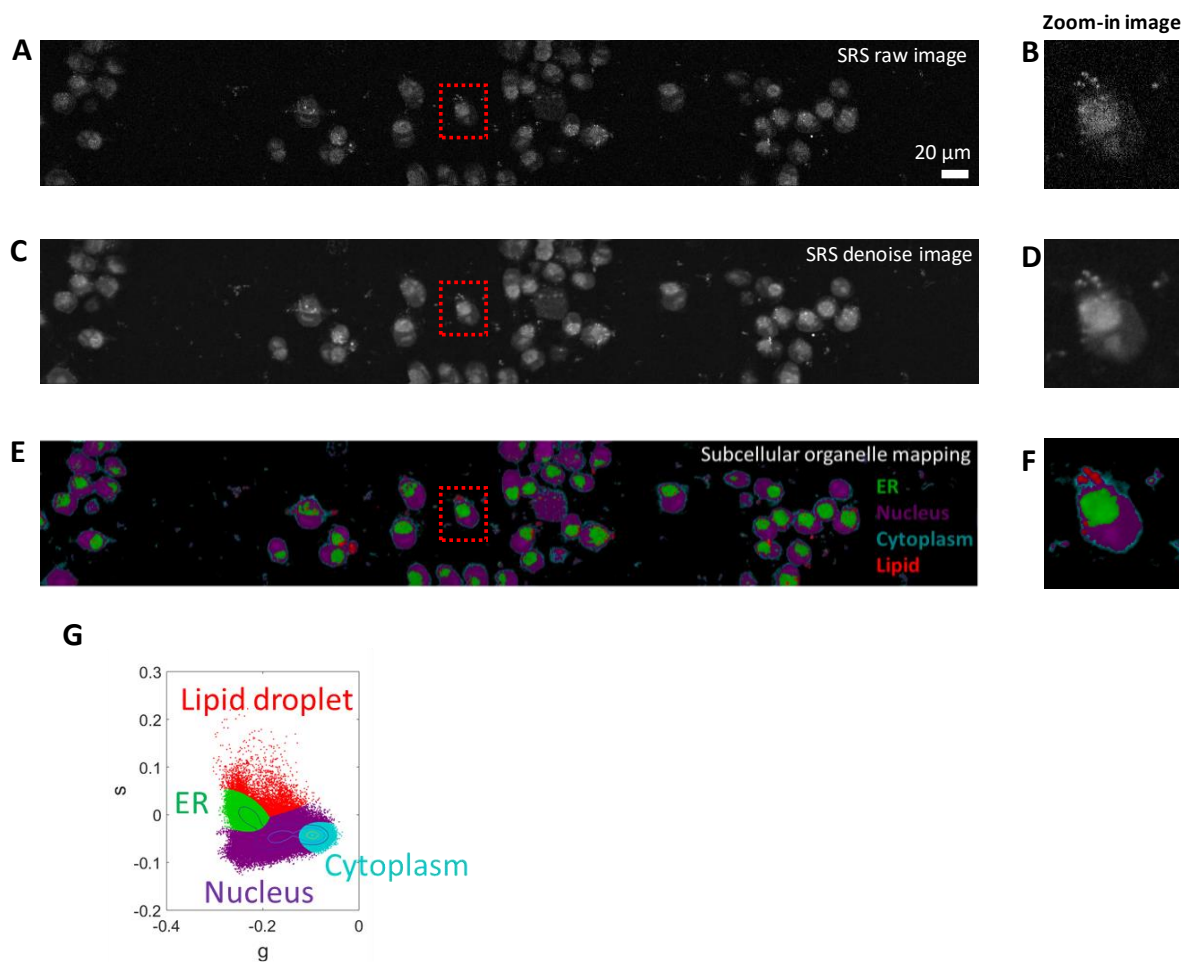

**Figure S2. Image denoise pre-processing for label-free chemical mapping.**
(A) SRS raw image at 2930 cm$^{-1}$ of living MIA PaCa-2 cells.
(B) Zoom-in area of the red dashed rectangular in (A).
(C) Denoised SRS image at 2930 cm$^{-1}$.
(D) Zoom-in area of the red dashed rectangular in (C).
(E) Sub-cellular organelle mapping results with the input from denoised SRS image (C).
(F) Zoom-in area of the red dashed rectangular in (E).
(G) Unsupervised clustered results in the 2-D phasor space, which is used to remap to the spatial domain and generate the chemical mapping in (E).

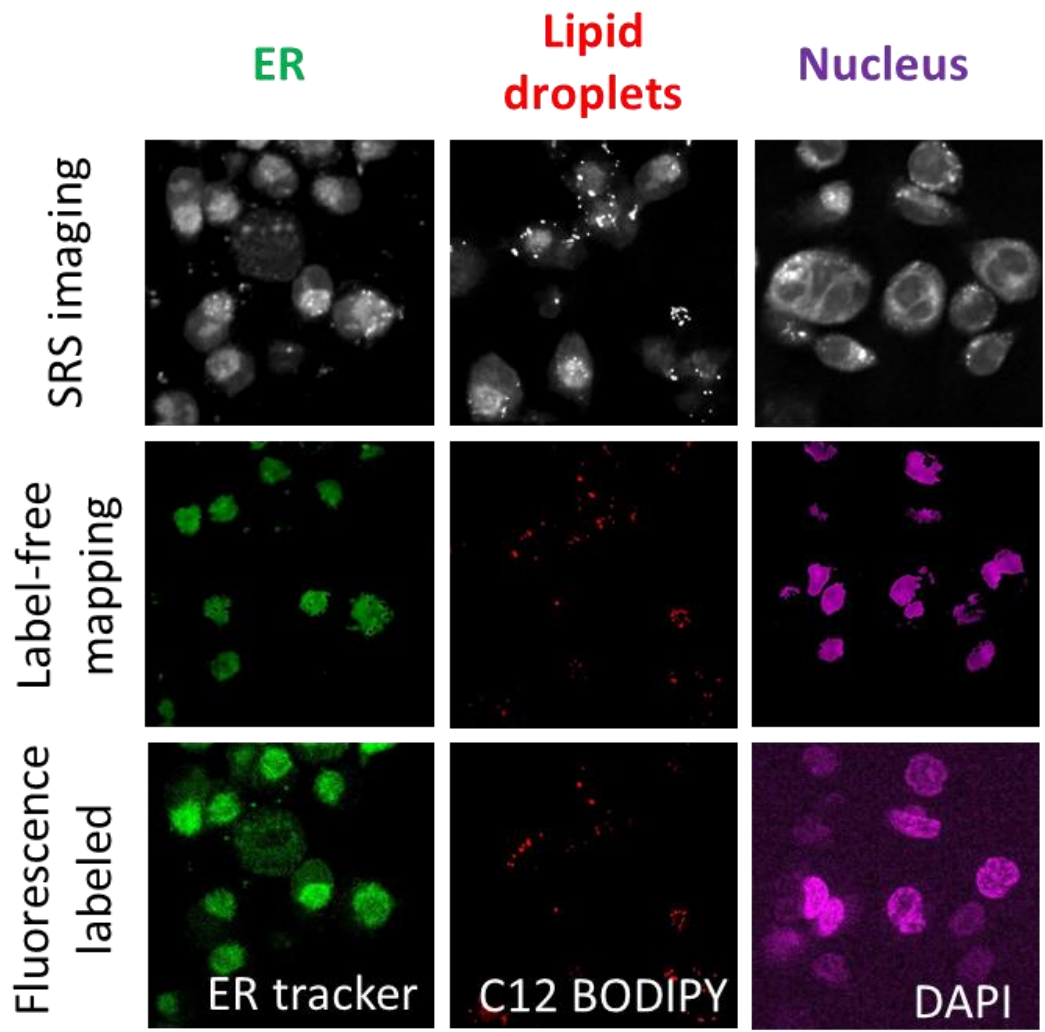

**Figure S3. Validation of sub-cellular organelles chemical mapping results with the fluorescence images.**

Upper: SRS images of MIA PaCa-2 cells (1st and 2nd row), and OVCAR-1 cells (3rd row). Middle: Label-free chemical maps results from multiplex SRS imaging cytometry. Lower: fluorescence images of ER labeled by ER tracker (left), lipid droplets labeled by C12 BODIPY (center), and nucleus labeled by DAPI (right).

| Cell | Lipid droplet out of ER | Lipid droplet within ER | ER | Nucleus | Cytoplasm |
|---|---|---|---|---|---|
| 'AreaShape_Area' | 'Mean_lipid_out_of_ER_AreaShape_Area' | 'Mean_lipid_within_ER_AreaShape_Area' | 'AreaShape_Area' | 'AreaShape_Area' | 'AreaShape_Area' |
| 'AreaShape_Compactness' | 'Mean_lipid_out_of_ER_AreaShape_Compactness' | 'Mean_lipid_within_ER_AreaShape_Compactness' | 'AreaShape_Compactness' | 'AreaShape_Compactness' | 'AreaShape_Compactness' |
| 'AreaShape_Eccentricity' | 'Mean_lipid_out_of_ER_AreaShape_Eccentricity' | 'Mean_lipid_within_ER_AreaShape_Eccentricity' | 'AreaShape_Eccentricity' | 'AreaShape_Eccentricity' | 'AreaShape_Eccentricity' |
| 'AreaShape_EulerNumber' | 'Mean_lipid_out_of_ER_AreaShape_EulerNumber' | 'Mean_lipid_within_ER_AreaShape_EulerNumber' | 'AreaShape_EulerNumber' | 'AreaShape_EulerNumber' | 'AreaShape_EulerNumber' |
| 'AreaShape_Extent' | 'Mean_lipid_out_of_ER_AreaShape_Extent' | 'Mean_lipid_within_ER_AreaShape_Extent' | 'AreaShape_Extent' | 'AreaShape_Extent' | 'AreaShape_Extent' |
| 'AreaShape_FormFactor' | 'Mean_lipid_out_of_ER_AreaShape_FormFactor' | 'Mean_lipid_within_ER_AreaShape_FormFactor' | 'AreaShape_FormFactor' | 'AreaShape_FormFactor' | 'AreaShape_FormFactor' |
| 'AreaShape_MajorAxisLength' | 'Mean_lipid_out_of_ER_AreaShape_MajorAxisLength' | 'Mean_lipid_within_ER_AreaShape_MajorAxisLength' | 'AreaShape_MajorAxisLength' | 'AreaShape_MajorAxisLength' | 'AreaShape_MajorAxisLength' |
| 'AreaShape_MaxFeretDiameter' | 'Mean_lipid_out_of_ER_AreaShape_MaxFeretDiameter' | 'Mean_lipid_within_ER_AreaShape_MaxFeretDiameter' | 'AreaShape_MaxFeretDiameter' | 'AreaShape_MaxFeretDiameter' | 'AreaShape_MaxFeretDiameter' |
| 'AreaShape_MaximumRadius' | 'Mean_lipid_out_of_ER_AreaShape_MaximumRadius' | 'Mean_lipid_within_ER_AreaShape_MaximumRadius' | 'AreaShape_MaximumRadius' | 'AreaShape_MaximumRadius' | 'AreaShape_MaximumRadius' |
| 'AreaShape_MeanRadius' | 'Mean_lipid_out_of_ER_AreaShape_MeanRadius' | 'Mean_lipid_within_ER_AreaShape_MeanRadius' | 'AreaShape_MeanRadius' | 'AreaShape_MeanRadius' | 'AreaShape_MeanRadius' |
| 'AreaShape_MedianRadius' | 'Mean_lipid_out_of_ER_AreaShape_MedianRadius' | 'Mean_lipid_within_ER_AreaShape_MedianRadius' | 'AreaShape_MedianRadius' | 'AreaShape_MedianRadius' | 'AreaShape_MedianRadius' |
| 'AreaShape_MinFeretDiameter' | 'Mean_lipid_out_of_ER_AreaShape_MinFeretDiameter' | 'Mean_lipid_within_ER_AreaShape_MinFeretDiameter' | 'AreaShape_MinFeretDiameter' | 'AreaShape_MinFeretDiameter' | 'AreaShape_MinFeretDiameter' |
| 'AreaShape_MinorAxisLength' | 'Mean_lipid_out_of_ER_AreaShape_MinorAxisLength' | 'Mean_lipid_within_ER_AreaShape_MinorAxisLength' | 'AreaShape_MinorAxisLength' | 'AreaShape_MinorAxisLength' | 'AreaShape_MinorAxisLength' |
| 'AreaShape_Orientation' | 'Mean_lipid_out_of_ER_AreaShape_Orientation' | 'Mean_lipid_within_ER_AreaShape_Orientation' | 'AreaShape_Orientation' | 'AreaShape_Orientation' | 'AreaShape_Orientation' |
| 'AreaShape_Perimeter' | 'Mean_lipid_out_of_ER_AreaShape_Perimeter' | 'Mean_lipid_within_ER_AreaShape_Perimeter' | 'AreaShape_Perimeter' | 'AreaShape_Perimeter' | 'AreaShape_Perimeter' |
| 'AreaShape_Solidity' | 'Mean_lipid_out_of_ER_AreaShape_Solidity' | 'Mean_lipid_within_ER_AreaShape_Solidity' | 'AreaShape_Solidity' | 'AreaShape_Solidity' | 'AreaShape_Solidity' |
| 'Children_MaskedER_Count' | 'Mean_lipid_out_of_ER_Distance_Centroid_Cells' | 'Mean_lipid_within_ER_Distance_Centroid_Cells' | 'Intensity_IntegratedIntensityEdge_Cell' | 'Intensity_IntegratedIntensityEdge_Cell' | 'Intensity_IntegratedIntensityEdge_Cell' |
| 'Children_MaskedNucleis_Count' | 'Mean_lipid_out_of_ER_Intensity_IntegratedIntensityEdge_Cell' | 'Mean_lipid_within_ER_Intensity_IntegratedIntensityEdge_Cell' | 'Intensity_IntegratedIntensity_Cell' | 'Intensity_IntegratedIntensity_Cell' | 'Intensity_IntegratedIntensity_Cell' |
| 'Children_lipid_out_of_ER_Count' | 'Mean_lipid_out_of_ER_Intensity_IntegratedIntensity_Cell' | 'Mean_lipid_within_ER_Intensity_IntegratedIntensity_Cell' | 'Intensity_LowerQuartileIntensity_Cell' | 'Intensity_LowerQuartileIntensity_Cell' | 'Intensity_LowerQuartileIntensity_Cell' |
| 'Children_lipid_within_ER_Count' | 'Mean_lipid_out_of_ER_Intensity_LowerQuartileIntensity_Cell' | 'Mean_lipid_within_ER_Intensity_LowerQuartileIntensity_Cell' | 'Intensity_MADIntensity_Cell' | 'Intensity_MADIntensity_Cell' | 'Intensity_MADIntensity_Cell' |
| 'Children_no_ER_Count' | 'Mean_lipid_out_of_ER_Intensity_MADIntensity_Cell' | 'Mean_lipid_within_ER_Intensity_MADIntensity_Cell' | 'Intensity_MassDisplacement_Cell' | 'Intensity_MassDisplacement_Cell' | 'Intensity_MassDisplacement_Cell' |
| 'Intensity_IntegratedIntensityEdge_Cell' | 'Mean_lipid_out_of_ER_Intensity_MassDisplacement_Cell' | 'Mean_lipid_within_ER_Intensity_MassDisplacement_Cell' | 'Intensity_MaxIntensityEdge_Cell' | 'Intensity_MaxIntensityEdge_Cell' | 'Intensity_MaxIntensityEdge_Cell' |
| 'Intensity_IntegratedIntensity_Cell' | 'Mean_lipid_out_of_ER_Intensity_MaxIntensityEdge_Cell' | 'Mean_lipid_within_ER_Intensity_MaxIntensityEdge_Cell' | 'Intensity_MaxIntensity_Cell' | 'Intensity_MaxIntensity_Cell' | 'Intensity_MaxIntensity_Cell' |
| 'Intensity_LowerQuartileIntensity_Cell' | 'Mean_lipid_out_of_ER_Intensity_MaxIntensity_Cell' | 'Mean_lipid_within_ER_Intensity_MaxIntensity_Cell' | 'Intensity_MeanIntensityEdge_Cell' | 'Intensity_MeanIntensityEdge_Cell' | 'Intensity_MeanIntensityEdge_Cell' |
| 'Intensity_MADIntensity_Cell' | 'Mean_lipid_out_of_ER_Intensity_MeanIntensityEdge_Cell' | 'Mean_lipid_within_ER_Intensity_MeanIntensityEdge_Cell' | 'Intensity_MeanIntensity_Cell' | 'Intensity_MeanIntensity_Cell' | 'Intensity_MeanIntensity_Cell' |
| 'Intensity_MassDisplacement_Cell' | 'Mean_lipid_out_of_ER_Intensity_MeanIntensity_Cell' | 'Mean_lipid_within_ER_Intensity_MeanIntensity_Cell' | 'Intensity_MedianIntensity_Cell' | 'Intensity_MedianIntensity_Cell' | 'Intensity_MedianIntensity_Cell' |
| 'Intensity_MaxIntensityEdge_Cell' | 'Mean_lipid_out_of_ER_Intensity_MedianIntensity_Cell' | 'Mean_lipid_within_ER_Intensity_MedianIntensity_Cell' | 'Intensity_MinIntensityEdge_Cell' | 'Intensity_MinIntensityEdge_Cell' | 'Intensity_MinIntensityEdge_Cell' |
| 'Intensity_MaxIntensity_Cell' | 'Mean_lipid_out_of_ER_Intensity_MinIntensityEdge_Cell' | 'Mean_lipid_within_ER_Intensity_MinIntensityEdge_Cell' | 'Intensity_MinIntensity_Cell' | 'Intensity_MinIntensity_Cell' | 'Intensity_MinIntensity_Cell' |
| 'Intensity_MeanIntensityEdge_Cell' | 'Mean_lipid_out_of_ER_Intensity_MinIntensity_Cell' | 'Mean_lipid_within_ER_Intensity_MinIntensity_Cell' | 'Intensity_StdIntensityEdge_Cell' | 'Intensity_StdIntensityEdge_Cell' | 'Intensity_StdIntensityEdge_Cell' |
| 'Intensity_MeanIntensity_Cell' | 'Mean_lipid_out_of_ER_Intensity_StdIntensityEdge_Cell' | 'Mean_lipid_within_ER_Intensity_StdIntensityEdge_Cell' | 'Intensity_StdIntensity_Cell' | 'Intensity_StdIntensity_Cell' | 'Intensity_StdIntensity_Cell' |
| 'Intensity_MedianIntensity_Cell' | 'Mean_lipid_out_of_ER_Intensity_StdIntensity_Cell' | 'Mean_lipid_within_ER_Intensity_StdIntensity_Cell' | 'Intensity_UpperQuartileIntensity_Cell' | 'Intensity_UpperQuartileIntensity_Cell' | 'Intensity_UpperQuartileIntensity_Cell' |
| 'Intensity_MinIntensityEdge_Cell' | 'Mean_lipid_out_of_ER_Intensity_UpperQuartileIntensity_Cell' | 'Mean_lipid_within_ER_Intensity_UpperQuartileIntensity_Cell' | 'Location_CenterMassIntensity_X_Cell' | 'Location_CenterMassIntensity_X_Cell' | 'Location_CenterMassIntensity_X_Cell' |
| 'Intensity_MinIntensity_Cell' | 'Mean_lipid_out_of_ER_Location_CenterMassIntensity_X_Cell' | 'Mean_lipid_within_ER_Location_CenterMassIntensity_X_Cell' | 'Location_CenterMassIntensity_Y_Cell' | 'Location_CenterMassIntensity_Y_Cell' | 'Location_CenterMassIntensity_Y_Cell' |
| 'Intensity_StdIntensityEdge_Cell' | 'Mean_lipid_out_of_ER_Location_CenterMassIntensity_Y_Cell' | 'Mean_lipid_within_ER_Location_CenterMassIntensity_Y_Cell' | 'Location_Center_X' | 'Location_Center_X' | 'Location_Center_X' |
| 'Intensity_StdIntensity_Cell' | 'Mean_lipid_out_of_ER_Location_Center_X' | 'Mean_lipid_within_ER_Location_Center_X' | 'Location_Center_Y' | 'Location_Center_Y' | 'Location_Center_Y' |
| 'Intensity_UpperQuartileIntensity_Cell' | 'Mean_lipid_out_of_ER_Location_Center_Y' | 'Mean_lipid_within_ER_Location_Center_Y' | 'Location_MaxIntensity_X_Cell' | 'Location_MaxIntensity_X_Cell' | 'Location_MaxIntensity_X_Cell' |
| 'Location_CenterMassIntensity_X_Cell' | 'Mean_lipid_out_of_ER_Location_MaxIntensity_X_Cell' | 'Mean_lipid_within_ER_Location_MaxIntensity_X_Cell' | 'Location_MaxIntensity_Y_Cell' | 'Location_MaxIntensity_Y_Cell' | 'Location_MaxIntensity_Y_Cell' |
| 'Location_CenterMassIntensity_Y_Cell' | 'Mean_lipid_out_of_ER_Location_MaxIntensity_Y_Cell' | 'Mean_lipid_within_ER_Location_MaxIntensity_Y_Cell' | 'Math_ER_cell_area_ratio' | 'Math_nucleus_cyto_mean_int_ratio' | 'SRS_phasor_coordinate_X' |
| 'Location_Center_X' | SRS_phasor_coordinate_X | SRS_phasor_coordinate_X | 'Math_ER_cell_integrateInt_ratio' | SRS_phasor_coordinate_X | SRS_phasor_coordinate_Y |
| 'Location_Center_Y' | SRS_phasor_coordinate_Y | SRS_phasor_coordinate_Y | SRS_phasor_coordinate_X | SRS_phasor_coordinate_Y | |
| 'Location_MaxIntensity_X_Cell' | | | SRS_phasor_coordinate_Y | | |
| 'Location_MaxIntensity_Y_Cell' | | | | | |
| 'Math_lipid_out_in_ER_count_ratio' | | | | | |
| 'Math_perimeter_area_ratio' | | | | | |
| 'Mean_MaskedER_Location_Center_X' | | | | | |
| 'Mean_MaskedER_Location_Center_Y' | | | | | |
| 'Mean_MaskedNucleis_Location_Center_X' | | | | | |
| 'Mean_MaskedNucleis_Location_Center_Y' | | | | | |
| 'RadialDistribution_FracAtD_Cell_1of4' | | | | | |
| 'RadialDistribution_FracAtD_Cell_2of4' | | | | | |
| 'RadialDistribution_FracAtD_Cell_3of4' | | | | | |
| 'RadialDistribution_FracAtD_Cell_4of4' | | | | | |
| 'RadialDistribution_MeanFrac_Cell_1of4' | | | | | |
| 'RadialDistribution_MeanFrac_Cell_2of4' | | | | | |
| 'RadialDistribution_MeanFrac_Cell_3of4' | | | | | |
| 'RadialDistribution_MeanFrac_Cell_4of4' | | | | | |
| 'RadialDistribution_RadialCV_Cell_1of4' | | | | | |
| 'RadialDistribution_RadialCV_Cell_2of4' | | | | | |
| 'RadialDistribution_RadialCV_Cell_3of4' | | | | | |
| 'RadialDistribution_RadialCV_Cell_4of4' | | | | | |

**Figure S4. List of 260 parameters extracted in single cell from multiplex SRS images by modules in *CellProfiler* to describe cellular and subcellular organelle including ER, lipid droplets within ER, lipid droplets out of ER, nucleus, and cytoplasm.**

These features were categorized into three groups including (1) the morphological features with green color background, (2) SRS intensity features with blue color background, and (3) SRS spectral features with orange color background.

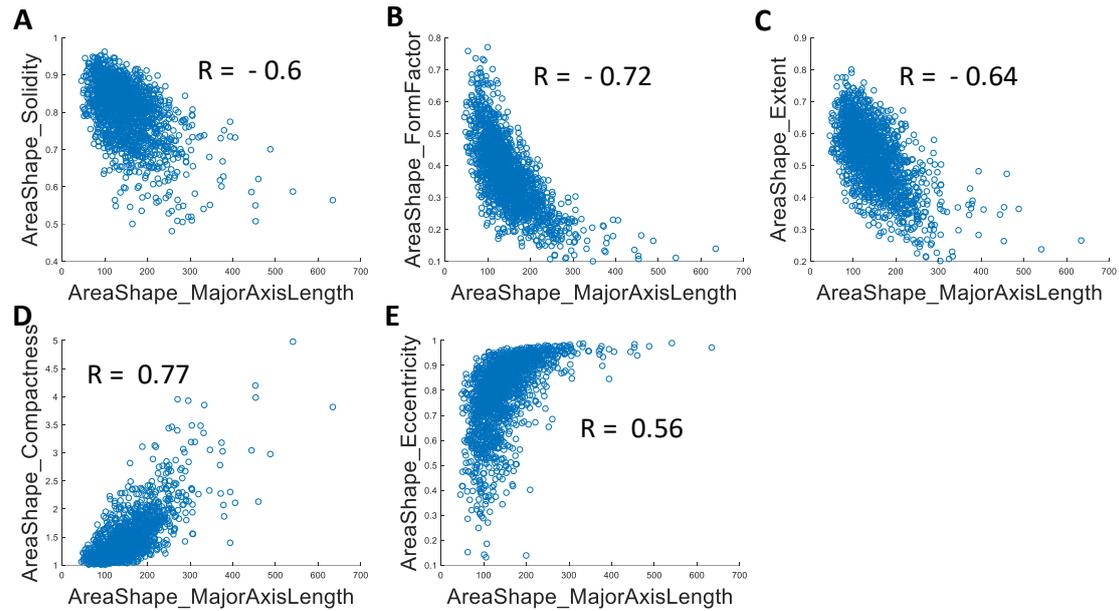

**Figure S5. The protrusion feature, the major axis length, correlates with the other cellular morphological protrusion features.**

(A) The cellular major axis length negatively correlates with the protrusion feature solidity with a correlation coefficient of -0.6. Solidity is defined as the proportion of the pixels in the convex hull that is also in the object, i.e., object area/convex hull area. Cell with protrusion structures tends to have lower solidity value.

(B) The cellular major axis length negatively correlates with the protrusion feature form factor with a correlation coefficient of -0.72. The form factor is calculated as $4*\pi*area/(perimeter^2)$. Cell with protrusion structure tends to have lower form factor value.

(C) The cellular major axis length negatively correlates with the protrusion feature extent with a correlation coefficient of -0.64. Extent is defined as the proportion of the pixels in the bounding box that is also in the region, i.e., object area/bounding box area. Cell with protrusion structure tends to have lower extent value.

(D) The cellular major axis length positively correlates with the protrusion feature compactness with a correlation coefficient of 0.77. Compactness is defined as the mean squared distance of the object's pixels from the centroid divided by the area. Cell with protrusion structure tends to have higher compactness value.

(E) The cellular major axis length positively correlates with the protrusion feature eccentricity with a correlation coefficient of 0.56. Eccentricity is defined as the ratio of the distance between the foci of the ellipse and its major axis length. Cell with protrusion structure tends to have higher eccentricity value.

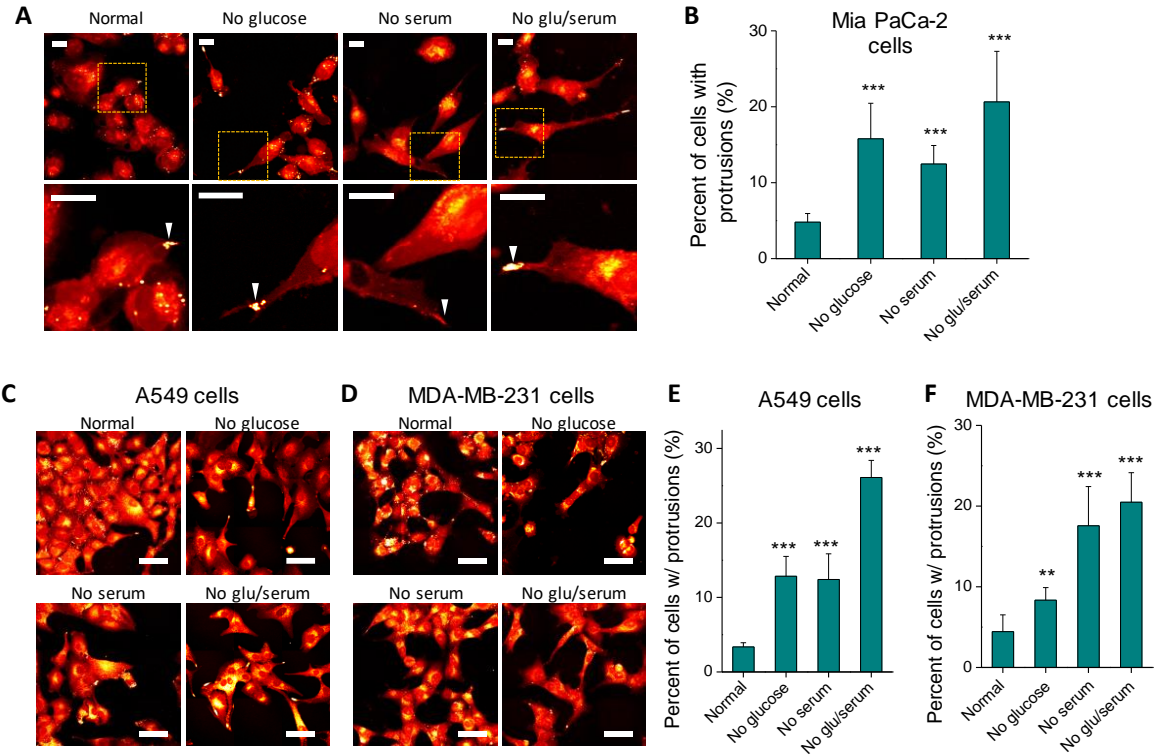

**Figure S6. LDs accumulation in protrusions augmented by starvation in MIA PaCa-2, A549, and MDA-MB-231 cells.**

(A) SRS images of MIA PaCa-2 cells in different culture media.
(B) Quantitation of the percentage of MIA PaCa-2 cells with protrusions in different culture media (n=5).
(C) SRS images of A549 cells in different culture media.
(D) SRS images of MDA-MB-231 cells in different culture media.
(E) Quantitation of the percentage of A549 cells with protrusions in different culture media (n=5).
(F) Quantitation on the percentage of MDA-MB-231 cells with protrusions in different culture media (n=5). The scale bars are 20 μm. ** $p < 0.01$, *** $p < 0.001$.

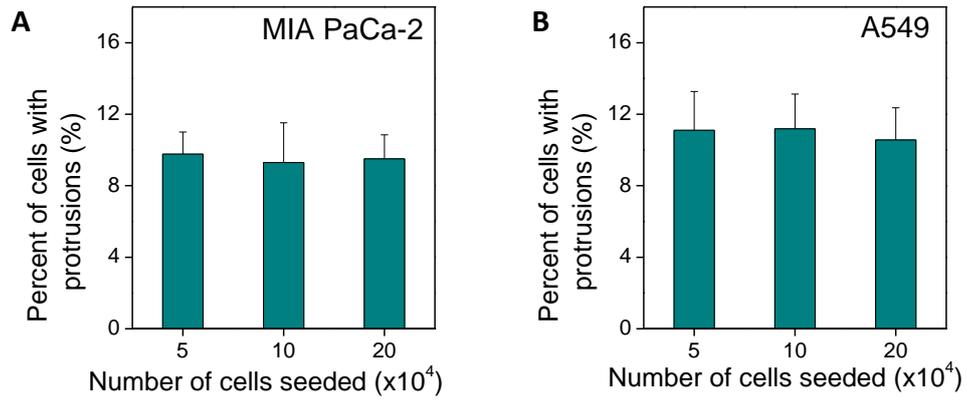

**Figure S7. Protrusion formation is independent of cell density.**

(A) The percentage of MIA PaCa-2 cells with protrusions as a function of cell concentration.
(B) The percentage of A549 cells with protrusions as a function of cell concentration. n=5.

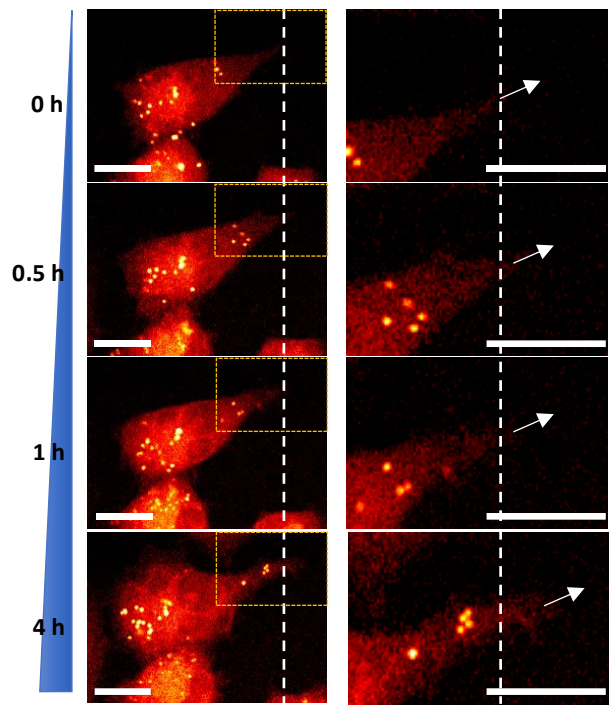

**Figure S8. Time dependent SRS images of MIA PaCa-2 cells under starvation.**

The dashed line marked the original boundary of the protrusion. The arrows point out the direction of the protrusion growth. Right panels are zoomed out images of the outlined regions in corresponding left panels. The scale bars are 20 μm.

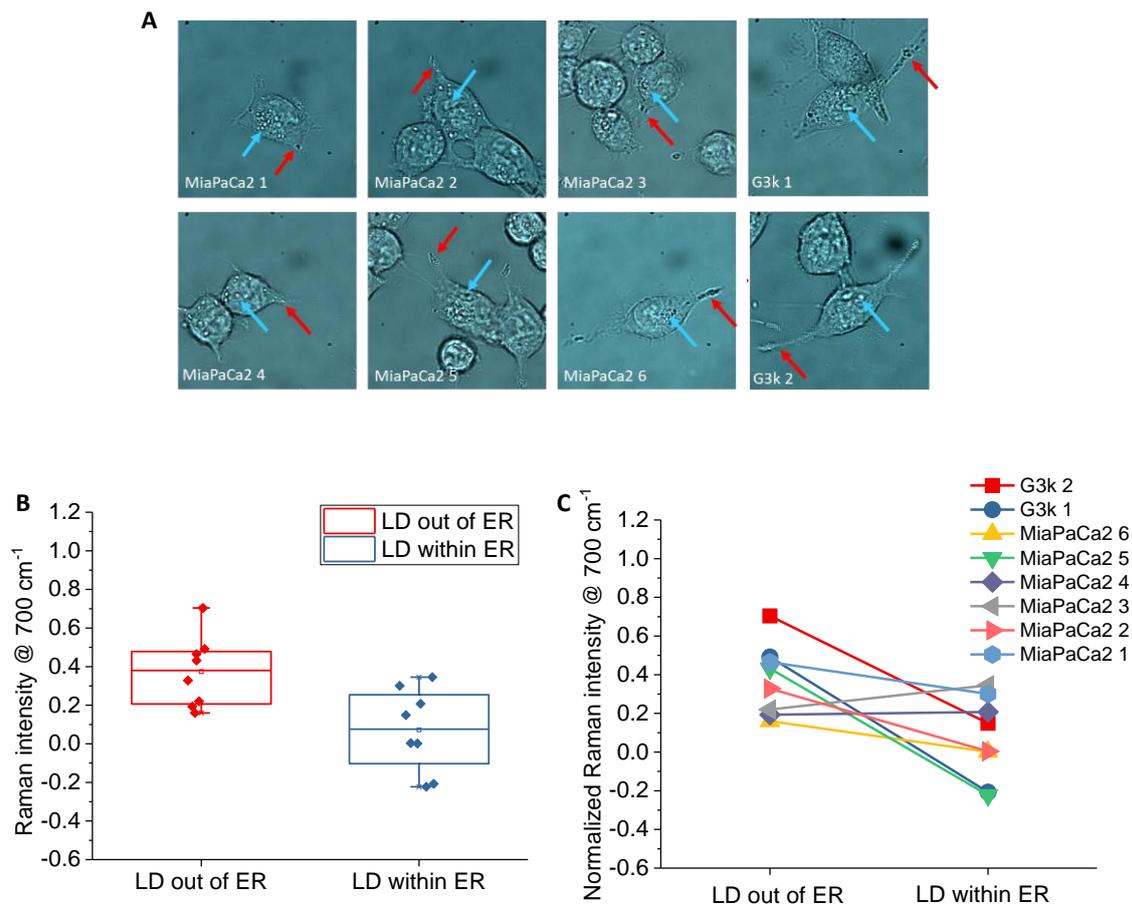

**Figure S9. Cholesterol-rich LDs out of ER statistically revealed by spontaneous Raman spectroscopy.**

(A) Transmission images of selected MIA PaCa-2 cells and G3K cells. The red arrows indicate the selected LD out of ER to probe, and the blue arrows indicate the selected LD within ER to probe. \
(B) Statistical analysis revealed that Raman intensity at 702 cm$^{-1}$ cholesterol signature peak of LD out of ER is higher than that of the LD within ER.
(C) Comparison of Raman intensity at 702 cm$^{-1}$ cholesterol signature peak between LDs out of ER and LDs within ER in the single cell.

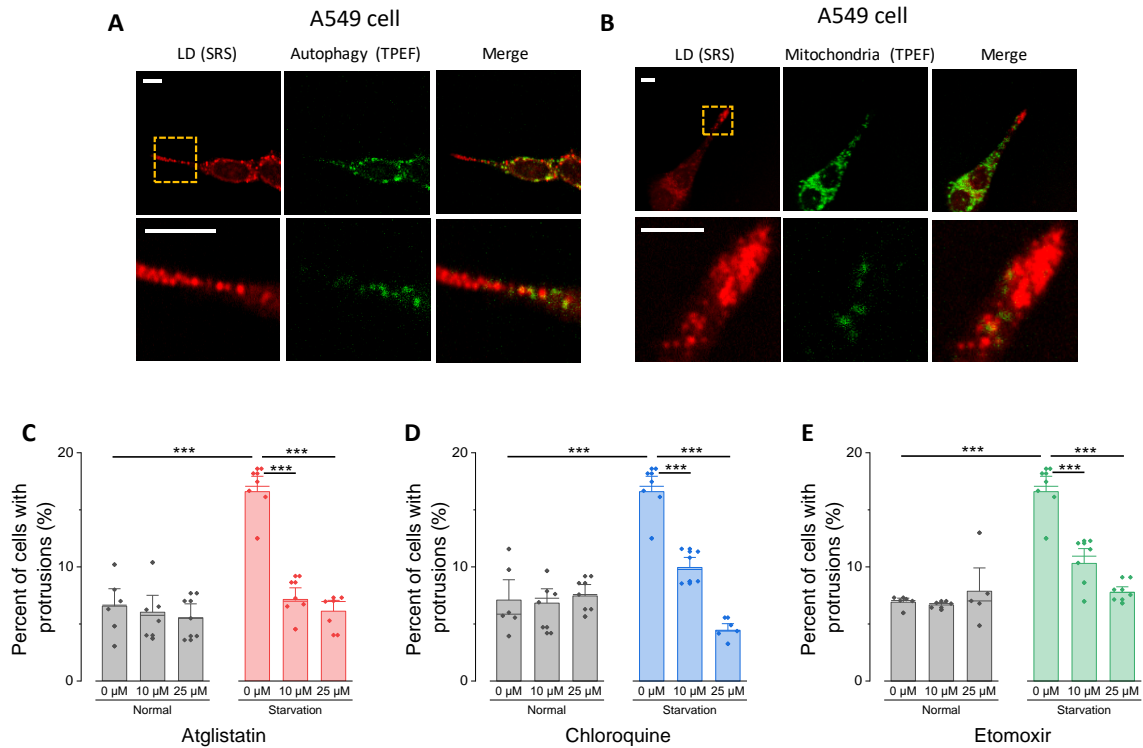

**Figure S10. LDs are degraded by autophagy and lipase and used in mitochondria for energy in A549 cell line.**

(A) SRS (left panel), TPEF (middle panel), and the composition (right panel) images from an A549 cell. TPEF imaging detects the autophagosomes labeled by monodansylcadaverine. The lower panels are the zoom-in areas (marked as dash square) shown in the upper panels.
(B) SRS (left panels), TPEF (middle panels), and the composition (right panels) images from an A549 cell labeled with MitoTracker™. The MitoTracker™ signal from mitochondria was acquired in TPEF imaging. The lower panels are the zoom-in areas (marked as dash square) shown in the upper panels.
(C) Percentage of cells with LD-rich protrusions after treatment by atglistation in normal and starvation conditions for A549 cells (n=5).
(D) Percentage of cell with LD-rich protrusions after treatment by chloroquine sulfate in normal and starvation conditions for A549 cells (n=5).
(E) Quantitation on the percentage of cells with LD-rich protrusions after treatment by etomoxir in normal and starvation conditions for A549 cells (n=5). The scale bars are 10 μm. *** $p < 0.001$.

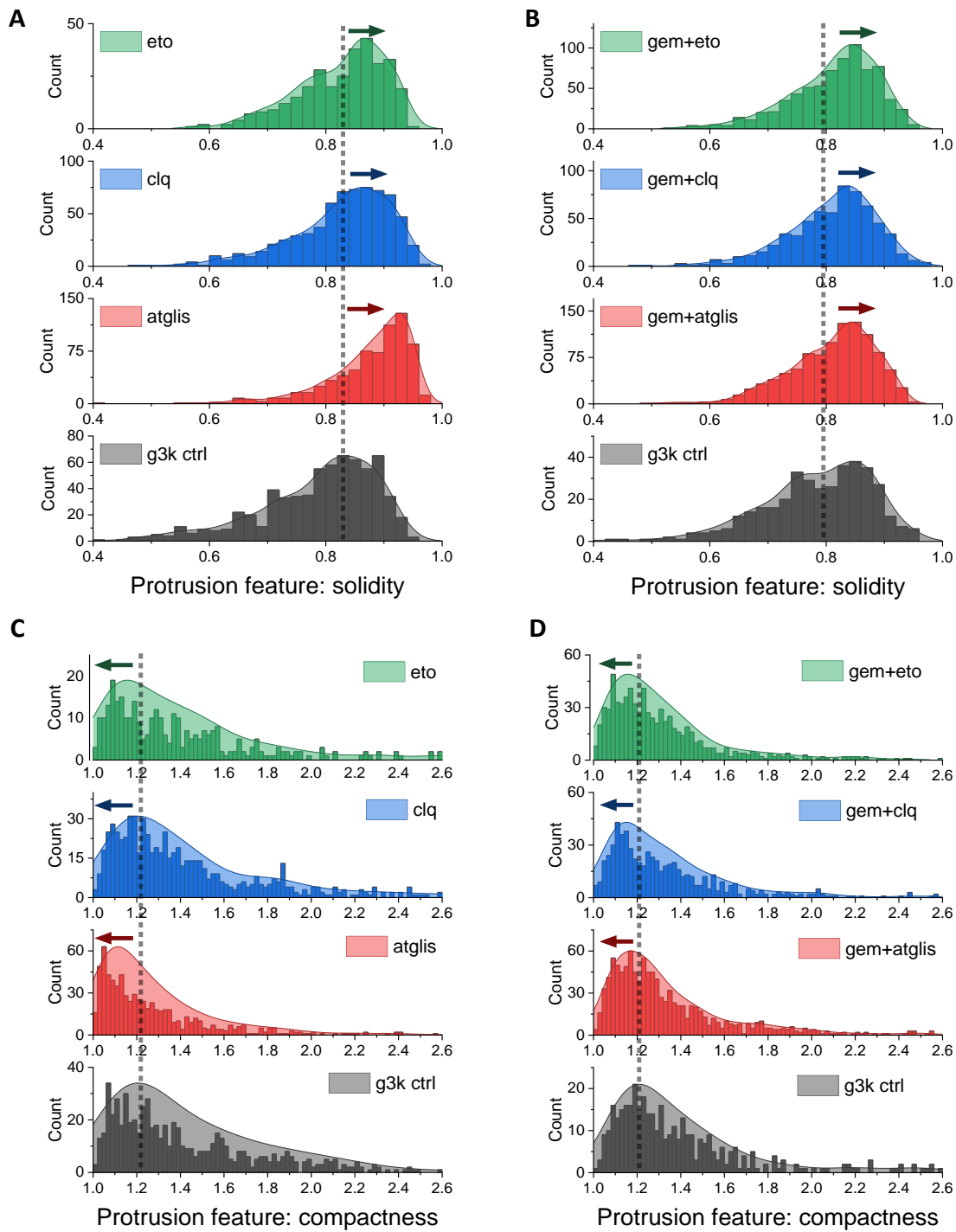

**Figure S11. Protrusion formation was suppressed with an ATGL inhibitor, an autophagy inhibitor, and a CPT1 inhibitor revealed by the other protrusion features solidity and compactness.**

(A and B) Histogram of solidity feature for (A), G3K cells and (B) gemcitabine treated G3K cells, treated with chloroquine sulfate, atglistatin, and etomoxir. The higher solidity value indicates cells with less protrusion formation.
(C and D), Histogram of compactness feature for (C) G3K cells and (D) gemcitabine treated G3K cells, treated with chloroquine sulfate, atglistatin, and etomoxir. The lower compactness value indicates cells with less protrusion formation.

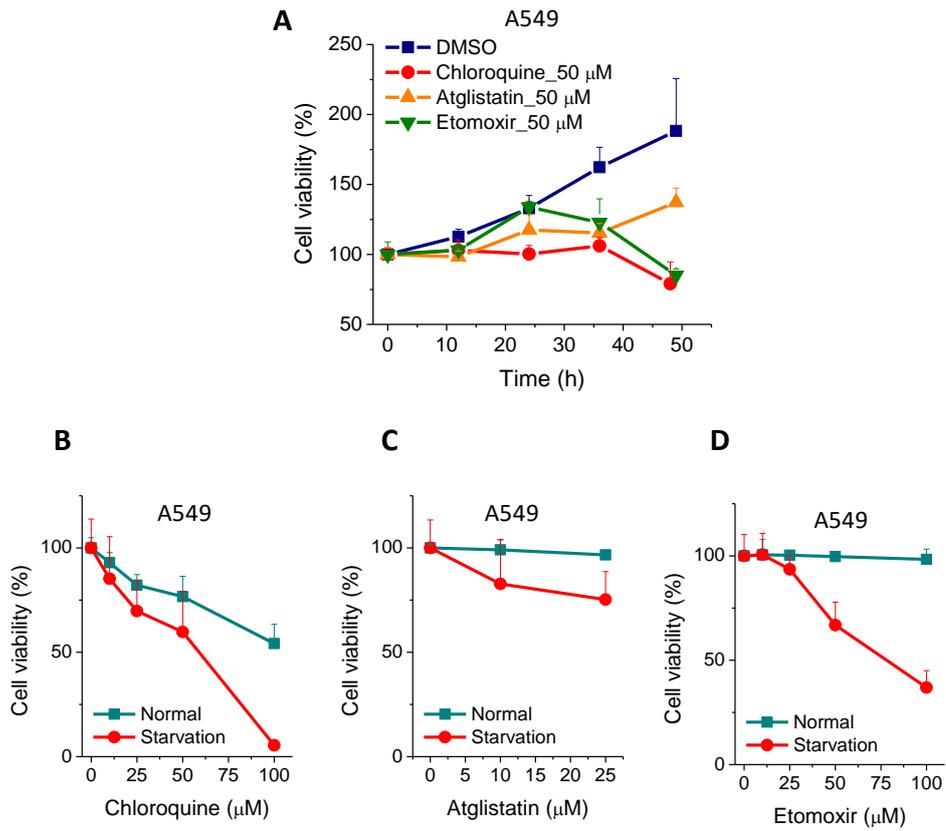

**Figure S12. Blockage of lipid metabolism suppressed cell survival under starvation in A549 cell line.**

(A) Time-dependent A549 cell viability with and without chloroquine sulfate treatment, atglistatin treatment, and etomoxir treatment under starvation condition.
(B-D) Concentration-dependent A549 cell viability when treated by, (B) chloroquine sulfate, (C) atglistatin, and (D) etomoxir under normal and starved conditions.

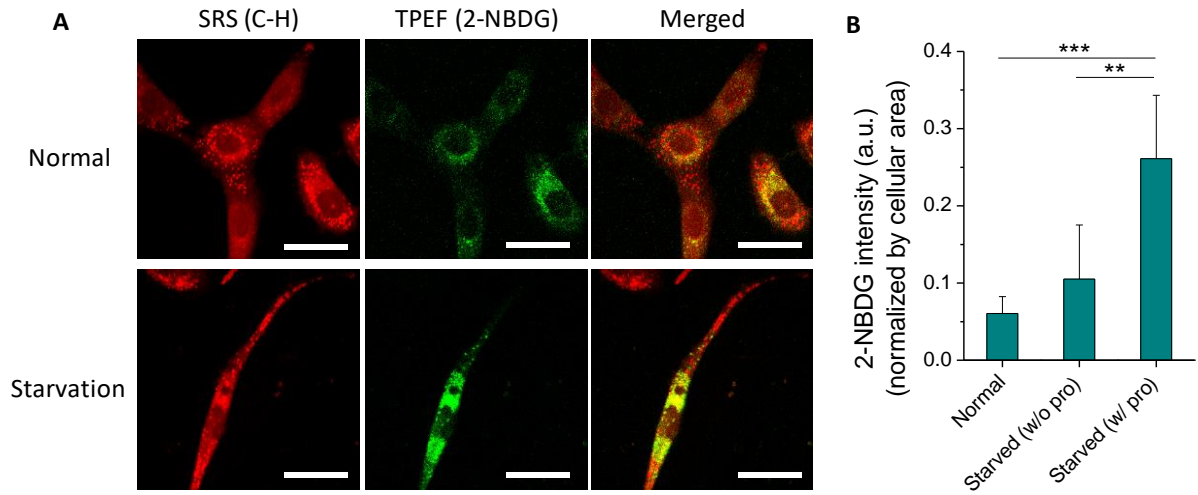

**Figure S13. Glucose analog 2-NBDG uptake in A549 cells.**

(A) SRS (left panels), TPEF (middle panels), and the composition (right panels) images from A549 cells incubated with 2-NBDG for 6 h. The TPEF signal is from the 2-NBDG accumulated in the cells. The upper panels are the control group, while the lower panels are collected under the starvation.
(B) Quantitation of the 2-NBDG intensity for A549 cells in normal and starvation conditions (with and without protrusions). The scale bars are 20 μm. n=5, ** $p < 0.01$, *** $p < 0.001$.

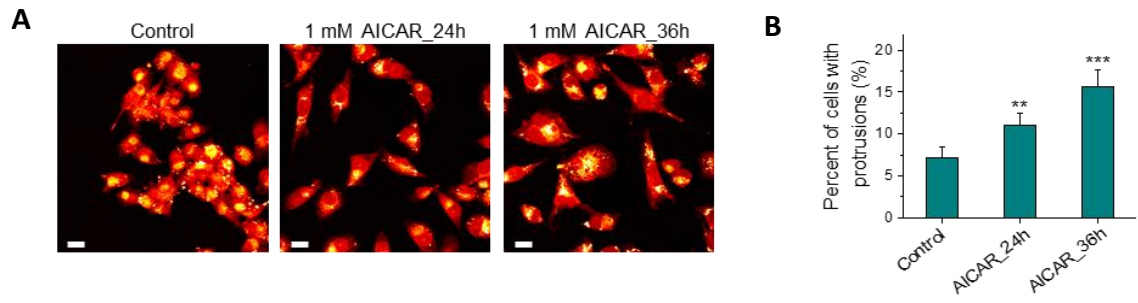

**Figure S14. LDs accumulation in cancer cell protrusions is mediated by AMPK activation.**

(A) SRS images of MIA PaCa-2 cells treated with 500 μM AICAR for different lengths of time.
(B) The percentage of cells with protrusions after AICAR treatment (n=5 for each condition). The scale bars are 10 μm. ** $p < 0.01$, *** $p < 0.001$.

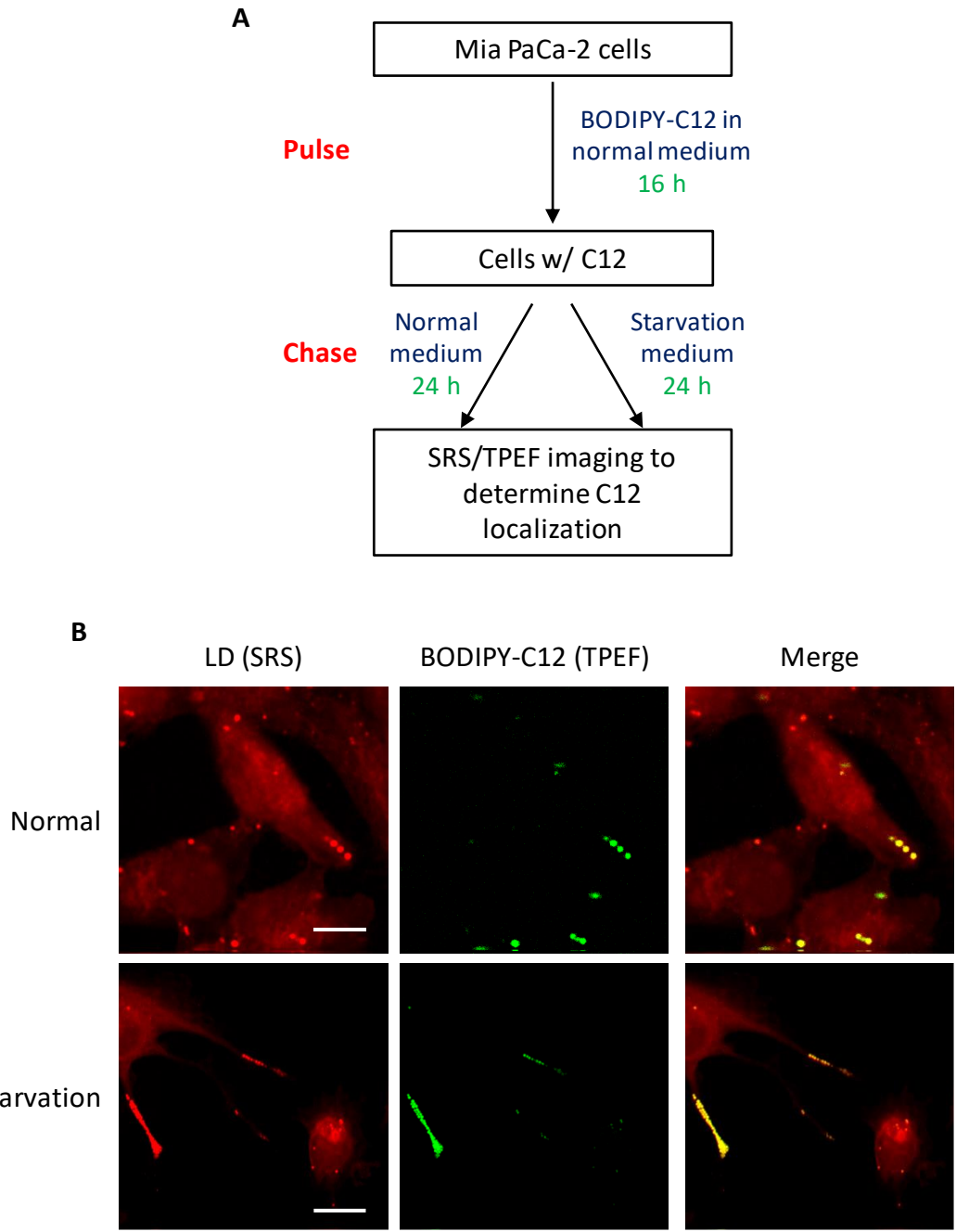

**Figure S15. BODIPY-C12 metabolism in MIA PaCa-2 cells revealed that fatty acids released from protrusion LDs are not used for new membrane synthesis.**

(A) Diagram of the pulse-chase experiment.
(B) SRS (left panels), TPEF (middle panels), and the composition (right panels) images from MIA PaCa-2 cells treated by BODIPY-C12 for 6 h. TPEF signal is from the BODIPY-C12 accumulated in cells. Upper panels are images acquired in normal condition, whereas the lower panels are images acquired under starvation. The scale bars are 10 μm.